\documentclass[aps,prd,preprintnumbers,superscriptaddress,linenumbers]{revtex4}
\usepackage{natbib}
\usepackage{amsmath}
\usepackage{amssymb}
\usepackage{amsthm}
\usepackage{todonotes}
\usepackage{mathtools}
\usepackage{mathrsfs}
\usepackage{bm}
\usepackage{slashed} 
\usepackage{graphicx}
\usepackage{multirow}
\usepackage{tikz}
\usepackage[caption=false]{subfig}
\usepackage{relsize}	 
\usepackage{array}
\usepackage{float}
\usepackage{color}
\usepackage{xcolor}
\usepackage{soul}
\usepackage{verbatim} 
\usepackage{hyperref}
\allowdisplaybreaks
\newcommand{\beq}{\begin{equation}}
	\newcommand{\eeq}{\end{equation}}
\newcommand{\be}{\begin{eqnarray}}
	\newcommand{\ee}{\end{eqnarray}}

\newcommand{\Q}{{\scriptsize Q}}
\newcommand{\G}{{\scriptsize G}}
\newcommand{\bsq}{{\boldsymbol q}^{\perp}}
\newcommand{\bsqq}{{q}^{\perp 2}}

\newcommand{\bsk}{{\boldsymbol k}^{\perp}}
\newcommand{\bska}{{\boldsymbol\kappa}^\perp}

\newcommand{\bskasq}{{\kappa}^{\perp2}}

\newcommand{\bsb}{{\boldsymbol b}^{\perp}}

\newcommand{\bsP}{{\boldsymbol P}^{\perp}} 
 
\newcommand{\bsp}{{\boldsymbol p}^{\perp}}

\newcommand{\es}{&=&}
\newcommand{\ps}{&+&}

\newcommand{\nn}{\nonumber}
\newcommand{\nnn}{\nonumber\\}

\begin{document}
	\title{Gluon contribution to the mechanical properties of a dressed quark in light-front Hamiltonian QCD}
	
	\author{Jai More}
	\email{jai.more@iitb.ac.in} \affiliation{ Department of Physics,
		Indian Institute of Technology Bombay, Powai, Mumbai 400076,
		India}
	
	\author{Asmita Mukherjee}
	\email{asmita@phy.iitb.ac.in} \affiliation{ Department of Physics,
		Indian Institute of Technology Bombay, Powai, Mumbai 400076,
		India}
	
	\author{Sreeraj Nair}
	\email{sreeraj@impcas.ac.cn} \affiliation{Institute of Modern Physics, Chinese Academy of Sciences, Lanzhou 730000, China}
	\affiliation{School of Nuclear Science and Technology, University of Chinese Academy of Sciences, Beijing 100049, China}
	\affiliation{CAS Key Laboratory of High Precision Nuclear Spectroscopy, Institute of Modern Physics, Chinese Academy of Sciences, Lanzhou 730000, China}
	
	\author{Sudeep Saha}
	\email{sudeepsaha@iitb.ac.in} \affiliation{ Department of Physics,
		Indian Institute of Technology Bombay, Powai, Mumbai 400076,
		India}

	\date{\today}
	
\begin{abstract}
We calculate the contribution to the gravitational form factors (GFFs) from the gluon part of the energy-momentum tensor in QCD. We take a simple spin $1/2$ composite state, namely a quark dressed with a gluon. We use the light-front Hamiltonian QCD approach in the light-front gauge. We also present the effect of the gluon on the mechanical properties like the pressure, shear and energy distributions of the dressed quark state. 

\end{abstract}

\maketitle
	
%
	
	\section{Introduction}
	The most fundamental question of hadron physics includes the origin of nucleon mass and nucleon spin structure.  The energy-momentum tensor (EMT) of QCD provides an understanding of these issues. 
	The  hadronic matrix element of the EMT sheds light on the sum rules and  gravitational coupling of quarks and gluons. This also gives an insight into the fundamental question of how the mass of the nucleon is formed from the quarks and  gluons \cite{Lorce:2021xku, Ji:1994av, Ji:1995sv, Lorce:2017xzd}. The coupling of gravitons and matter particles is through gravitational form factors (GFFs) \cite{Polyakov:2018zvc}. These GFFs are the form factors of EMT of QCD. They are analogues to electromagnetic form factors in QED \cite{Pagels:1966zza}.
 
  GFFs can be related to generalized parton distributions (GPDs) \cite{Polyakov_2003}, which are accessible via high-energy scattering processes like deeply virtual Compton scattering (DVCS) at experimental facilities like Jefferson Lab (JLab)  \cite{dHose2016,Kumeri_ki_2016} and deeply virtual meson production (DVMP) \cite{Collins_1997}.  The Ji's sum rule relates the nucleon GFFs $A(q^2)$ and $B(q^2)$ to the angular momentum  carried by the quarks \cite{PhysRevLett.78.610}. The quark GFFs for the nucleon were studied at JLab from DVCS \cite{CLAS:2007clm}.  Some of the experimental facilities that will further improve the constraints on the quark GFFs via the GPDs are the upcoming Electron-ion collider (EIC) at BNL \cite{AbdulKhalek:2021gbh}, International Linear Collider (ILC), the Japan proton accelerator complex (J-PARC) \cite{Kumano:2022cje}, the nuclotron-based ion collider facility (NICA) \cite{MPD:2022qhn} and the PANDA experiment at the Facility for Antiproton and Ion Research (FAIR) \cite{PANDA:2009yku}.
 	
  The parameterization of the EMT matrix element in terms of different hadrons for different spins can be considered as shown in Ref. \cite{Cosyn:2019aio, Polyakov:2019lbq, Cotogno:2019vjb, Panteleeva:2020ejw, Kim:2020lrs}. For a spin-half system like a proton, the matrix elements of the symmetric EMT can be parametrized in terms of four GFFs.  
	  The GFFs are functions of the square of the momentum transfer $(q^2)$ in the process.  The form factor $A(q^2)$ for hadrons of any spin is constrained by the conservation of momentum, such that $A(0)$ when summed over all partons at zero-momentum transfer is unity. Likewise, the GFF $B(q^2)$ is also constrained at zero-momentum transfer to be zero. For a spin-half fermion, the constraint on $B(q^2)$ is also referred to as the vanishing of the ``anomalous gravitomagnetic moment" in analogy to the anomalous magnetic moment \cite{Polyakov:2018zvc}. The GFF $\overline{C}(q^2)$ can be non-zero for quarks and gluons separately, due to the non-conservation of the partial EMT; however, it is expected to vanish when summed over all quarks and gluons  \cite{Lorce:2018egm}. 
	Unlike other GFFs, the ``\emph{D}-term"  or the {\it Druck term} is unconstrained at zero-momentum transfer and can only be determined experimentally. The GFFs can give an insight into the basic mechanical properties of the nucleons like the mass, spin and pressure distributions. In fact, by studying the form factor $D(q^2)$, one can get information about the pressure and shear distributions inside the proton \cite{Polyakov:2018zvc}.	Recent extraction of the $D$-term from JLab data has allowed visualizing the pressure and shear force distributions \cite{Burkert:2018bqq, Burkert:2021ith}. The pressure distribution is repulsive near the core of the nucleon and attractive towards the outer region. In fact near the centre, the magnitude of the pressure is comparable to the pressure distribution inside a neutron star, which is the densest object in the universe. The fact that the pressure and energy distributions can be obtained from the GFFs which are accessible through the generalized parton distributions, suggests an  interesting method  to study the equation of state of dense matters like neutron star \cite{Rajan:2018zzy}. The pressure, shear and energy distributions are frame dependent. 
   A discussion of the definitions of these distributions in  different reference frames can be seen in \cite{Lorce:2018egm}. In many cases, one defines such distributions in the Breit frame \cite{Polyakov:2018zvc}. However, in this frame, they are subject to relativistic corrections. In \cite{Lorce:2018egm} and later in \cite{Freese:2021czn} two-dimensional light-front distributions are introduced. A comparative discussion of the definitions of densities in different frames can be found in \cite{Freese:2022fat}. Because of the Galilean symmetry in the light-front framework, such 2D distributions are fully relativistic. A connection between the 2D and 3D distributions can be obtained in terms of Abel transformation \cite{Panteleeva:2021iip, Choudhary:2022den}. 
 The gravitational form factors have been extensively studied for the nucleon with various models namely the simple multi-pole model \cite{Lorce:2018egm}, the chiral quark soliton model \cite{Goeke:2007fp,Wakamatsu:2007uc,Schweitzer:2002nm}, the Bag model \cite{Neubelt:2019sou}, the Skyrme model \cite{Cebulla_2007,Kim_2012}, AdS/QCD motivated diquark model \cite{Chakrabarti:2015lba}, in chiral perturbation theory \cite{Dorati_2008,Chen_2002,Belitsky_2002}, and in Lattice QCD \cite{Hagler:2003jd,Gockeler:2003jfa,LHPC:2010jcs,QCDSF-UKQCD:2007gdl,Deka:2013zha}. Recently \emph{D}-term of the nucleon in a holographic QCD model has been calculated in \cite{Fujita:2022jus}. 
 
In most of the phenomenological  models for the nucleon, gluonic degrees of freedom are not included, as a result, only the quark GFFs can be perceived. However, some of the GFFs like the \emph{D}-term, that contributes to the pressure and shear force distributions in the nucleon depend on the so-called `bad' components of the energy-momentum tensor, that includes quark-gluon interactions. Thus it is important to investigate the role played by the gluons in such distributions.  However, the gluon GFFs are much less studied  theoretically and  fewer constraints exist for them from experiments. Gluon studies for the nucleon include the calculation of the gluon $A(q^2)$  in an extended holographic light-front QCD framework\cite{deTeramond:2021lxc},  GFFs for the gluon have also been calculated recently in Lattice QCD \cite{Shanahan_2019}. The gluon GPDs and form factors have been calculated in a soft-wall ADS/QCD model in \cite{Lyubovitskij:2020xqj}. In a very recent paper \cite{Tan:2023kbl}, gluon GPDs have been  calculated in a light-front spectator model, where the proton state is assumed to be consisting of one gluon and one spectator particle containing three valence quarks.  The gluon GPDs and GFFs for nucleon have been investigated in a holographic QCD framework in the context of photoproduction or leptoproduction of $J/\Psi$ and $\Upsilon$ in \cite{Mamo:2019mka}. Such experiments are possible at the JLab \cite{GlueX:2019mkq}. The future EIC will  focus on the study of nucleon structure and specifically, it would seek extraction of gluon \emph{D}-term for the first time \cite{AbdulKhalek:2021gbh, Anderle:2021wcy}.

 In this work, we study the gluon GFFs in a field theoretical model of a relativistic spin half system, namely a quark dressed with a gluon at one loop. We use the light-front Hamiltonian approach, in which the dressed quark state can be expanded in Fock space in terms of multi-parton light-front wave functions (LFWFs). The two-particle quark-gluon LFWF can be obtained analytically from the light-front QCD Hamiltonian.  We employ the resulting two-particle light-front wave functions to calculate the necessary overlap expressions using two-component representation \cite{Zhang:1993dd}. A similar model and approach have been used earlier to investigate the GPDs and Wigner functions   \cite{Chakrabarti:2004ci,Chakrabarti:2005zm,Brodsky:2006in,Brodsky:2006ku,Mukherjee:2014nya,Mukherjee:2015aja,More:2017zqq,More:2017zqp}.  The advantage is that in the light-front gauge, $A^+=0$, one can eliminate the constrained degrees of freedom which then allows an analytic calculation of the matrix element of  all the components of the EMT for such state. Thus one can explore the effect of the quark-gluon interaction in QCD that plays a major role  for example in the \emph{D}-term. The quark GFFs and the mechanical properties were investigated in this model in an earlier publication \cite{More:2021stk}. Here, we investigate the GFFs from the gluon part of the EMT as well as the contribution to the two-dimensional pressure, shear and energy distributions coming from the gluon.
The manuscript is organized in the following manner: In Sec II, we discuss the two-component formalism and present the calculation of the gravitational form factors for gluon in the light-front dressed quark state. In Sec III, using \emph{D}-term we calculate the pressure distributions, forces and energy densities. 
 In Sec IV we summarize our results. The useful formulas and essential steps of calculations can be found in the appendices. 
\section{Gravitational form factors and the two-component formalism}

	 In this section, we discuss the method used to obtain the gravitational form factors (GFFs) using a dressed quark state in light-front Hamiltonian QCD.
\\
	The symmetric QCD EMT is defined as, 
	\be
	\theta^{\mu \nu} \es \theta^{\mu \nu}_{\Q}+\theta^{\mu \nu}_\G,\\
	\theta^{\mu \nu}_{\Q}\es 
	\frac{1}{2}\overline{\psi}\ i\left[\gamma^{\mu}D^{\nu}+\gamma^{\nu}D^{\mu}\right]\psi - g^{\mu \nu} \overline{\psi} \left(
	i\gamma^{\lambda}D_{\lambda} -m
	\right)
	\psi \label{emtqcd},\\
	\theta^{\mu \nu}_\G\es - F^{\mu \lambda a}F_{\lambda a}^{\nu} + \frac{1}{4} g^{\mu \nu} \left( F_{\lambda \sigma a}\right)^2 .
	\ee
	
	  The last term in Eq.~\ref{emtqcd} will become zero because of the equation of motion.
   	A standard way to  parameterize the matrix element in terms of the  EMT for a spin-1/2 system is given by
	\be
	\langle P^{\prime}, S^{\prime}|	\theta^{\mu\nu}_i(0)|P,S \rangle \es\overline{U} (P^{\prime}, S^{\prime})\bigg[-B_i(q^2)\frac{\overline{P}^{\mu}\ \overline{P}^{ \nu}}{m}+\left(A_i(q^2)+B_i(q^2)\right)\frac{1}{2}(\gamma^{\mu}\overline{P}^{\nu}+\gamma^{\nu}\overline{P}^{\mu})\nnn
	\ps C_i(q^2)\frac{q^{\mu}q^{\nu}-q^2g^{\mu\nu}}{m}+\overline{C}_i(q^2)m\ g^{\mu\nu}\bigg]U(P, S), 
	\label{FF}
	\ee
	where   the Lorentz indices $(\mu, \nu)\ \equiv \{+,-,1,2\}$, $\overline{P}^{\mu}=\frac{1}{2}(P^{\prime}+P)^{\mu}$ is the average nucleon four-momentum. $\overline{U}(P', S'), U(P, S)$ are the Dirac spinors for the state, and $m$ is the mass of the target state , $i\equiv(Q, G)$. $A_i, B_i, C_i $ and  $\overline{C_i}$ are the quark or gluon gravitational form factors. One also uses the notation $D(q^2) = 4 C(q^2)$. As stated in the introduction, $A(q^2)$, $B(q^2)$ and $\overline{C}(q^2) $  are constrained, but $D(q^2)$ is unconstrained \cite{Lorce:2018egm}.  In general, from the matrix element of the energy-momentum tensor, the GFFs can be extracted. GFFs give how matter couples with gravity.
In this work, we study the gluon GFFs and analyze their contribution to the  mechanical properties of a quark state dressed by a gluon by extracting them from the gluon part of the EMT. This builds upon our previous study \cite{More:2021stk}, in which we used the quark part of the EMT to extract the quark GFFs and analyzed their contribution to the mechanical properties.
 
	
	In light front Hamiltonian formalism, a state with momentum $P$ and helicity $\lambda$ can be expanded in Fock space in terms of the light front wave functions (LFWFs). The LFWFs are boost invariant.  Here we consider a quark state dressed with one gluon, that is, we truncate the Fock space expansion at the two-particle level. The state can be written as 
	\be \label{state}
	|P,\lambda \rangle 
	\es \psi_1 (P, \lambda) b^{\dagger}_{\lambda}(P)|0 \rangle + \sum_{\lambda_1, \lambda_2}\int[k_1][k_2]   \sqrt{2(2\pi)^3P^+} \delta^3 (P-k_1-k_2)\ \psi_2 (P,\lambda|k_1,\lambda_1;k_2,\lambda_2) b^{\dagger}_{\lambda_1}(k_1) a^{\dagger}_{\lambda_2}(k_2)|0\rangle,  \nnn 
	\text{where}~~ [k]\es \frac{dk^+ d^2\bsk} {\sqrt{2(2\pi)^3{k^{+}}}}.
   \label{state}
	\ee
	In Eq. \ref{state}, $\psi_1(P, \lambda)$ in the first term, corresponds to a single particle with momentum (helicity) $P (\lambda)$ and also gives the normalization of the state. The two-particle LFWF, $\psi_2(P,\lambda|k_1,\lambda_1;k_2,\lambda_2)$ is related to the probability amplitude of finding two particles namely a quark and a gluon with momentum (helicity) $k_1(\lambda_1)$ and $k_2(\lambda_2)$, respectively,  inside the dressed quark state. 
	$b^\dagger$ and $a^\dagger$ correspond to the creation operator of quark and gluon respectively. 
	
	The LFWFs can be written in terms of relative momenta so that they are independent of the momentum of the composite  state. The relative momenta $x_i$, $\bska_i$ are defined  such that they satisfy the relation $x_1+x_2=1$ and $\bska_1+\bska_2=0$.
	\be
	k_i^+=x_iP^+,~~~~ \bsk_i=\bska_i+x_i \bsP,
	\ee 
	where $x_i$ is the longitudinal momentum fraction for the quark or gluon, inside the two-particle LFWF.
	The boost invariant two-particle LFWF can be written as,
	\be\label{BILFWF}
	\phi^{\lambda a}_{\lambda_1,\lambda_2}(x,\bska)\es \frac{g}{\sqrt{2(2\pi)^3}}
\bigg[\frac{x(1-x)}{\bskasq+m^2x^2}\bigg]\frac{T^a}{\sqrt{x}} \chi_{\lambda_1}^{\dagger}\nnn
	&\times&\bigg[\frac{2(\bska\cdot \epsilon_{\lambda_2}^{\perp*})}{x}+\frac{1}{1-x}(\tilde{\sigma}^{\perp}\cdot\bska)(\tilde{\sigma}^{\perp}\cdot \epsilon_{\lambda_2}^{\perp*})+im(\tilde{\sigma}^{\perp}\cdot \epsilon_{\lambda_2}^{\perp*})\frac{x}{1-x}\bigg]\chi_{\lambda} \psi_1^{\lambda},
	\ee
	where, $\phi^{\lambda a}_{\lambda_1,\lambda_2}(x_i,\bska_i)=\sqrt{P^+}\psi_2 (P,\lambda|k_1,\lambda_1;k_2,\lambda_2)$,  $g$ is the quark-gluon  coupling. $T^a$ and $\boldsymbol{ \epsilon}_{\lambda_2}^\perp$ are colour SU(3) matrices and polarization vector of the gluon. The quark mass and the two-component spinor for the quark are denoted by $m$ and $\chi_\lambda$ respectively, $\lambda=1,2$ correspond to helicity up/down. We have used the notation $\tilde{\sigma}_1=\sigma_2$ and $\tilde{\sigma}_2=-\sigma_1$ \cite{Harindranath:2001rc}. Note that, here $x$ and $\bska$ is longitudinal momentum fraction and the relative transverse momentum of gluon respectively. We have used the  two-component framework developed in light-cone gauge $A^+=0$  \cite{Zhang:1993dd}. In this gauge, by using a suitable  representation of the gamma matrices one can write:
	\begin{align}
		\psi_+= \begin{bmatrix}
			\xi\\0
		\end{bmatrix}, ~~~~\psi_-=\begin{bmatrix}
			0\\ \eta
		\end{bmatrix}, 
	\end{align}
	where the two-component quark fields are given by 
	\be
	\xi(y) \es \sum_{\lambda}\chi_{\lambda}\int \frac{[k]}{\sqrt{2(2\pi)^3}}[b_{\lambda}(k)e^{-ik\cdot y}+d^{\dagger}_{-\lambda}(k)e^{ik\cdot y}],
	\\
	\eta(y) \es \left(\frac{1}{i\partial^+}\right)\left[\sigma^{\perp}\cdot\left(i\partial^{\perp}+g A^{\perp}(y)\right)+im\right]\xi(y),
	\ee
	$\eta(y)$ is the constrained field, which may be eliminated using the above equation. The dynamical components of the gluon field are given by 
	\begin{align}
		A^{\perp}(y)= \sum_{\lambda} \int \frac{[k]}{\sqrt{2(2\pi)^3k^+}}[{\bf\epsilon}^{\perp}_{\lambda}a_{\lambda}(k)e^{-i k \cdot y}+ {\bf \epsilon}^{\perp*}_{\lambda}a^{\dagger}_{\lambda}(k)e^{i k \cdot y}].
	\end{align}
	Here we have suppressed the colour indices.
	The four momenta in light-front coordinates are defined as 
	\be
	P^\mu \es(P^+,\bsP,P^-). 
	\ee
	 We choose a frame where the four momenta of the initial and the final state are given by : 
	\be\label{initialmom}
	P^{\mu} \es\bigg(P^+, {\bf0}^{\perp}, \ \frac{m^2}{P^+}\bigg),\\
	\label{finalmom}
	P^{\prime\mu}\es\bigg(P^+,\ \bsq,\ \frac{\bsqq  + m^2}{P^+}\bigg), 
	\ee
	and the invariant momentum transfer
	\be\label{momtranfer}
	q^\mu\es(P^{\prime}-P)^\mu=\bigg(0, \ \bsq,   \frac{\bsqq}{P^+}\bigg).
	\ee
 This is the Drell-Yan frame (DYF), in this frame, the momentum transfer is purely in the transverse direction, and $q^+=0$, in other words $q^2=-\bsqq$.

 In order to calculate the GFFs, we define the matrix element of the EMT as follows
	\be \label{matrixelement}
	\mathcal{M}^{\mu \nu }_{SS'} = \frac{1}{2}\left[\langle P'  ,S'|	\theta^{\mu \nu }_\G(0)|P,S \rangle \right],
	\ee
	where 
 $(S, S') \equiv \{ \uparrow,\downarrow \}$  is the helicity of the initial and final state. $\uparrow(\downarrow)$ positive (negative) spin projection along $z-$ axis. We use the dressed
 quark state in Eq. (\ref{state}) to calculate the matrix element.
 The form factor $A_\G(q^2)$ and $B_\G(q^2)$ can  be calculated from the `good' components ($\theta_\G^{++}$) directly by taking suitable combinations. 
Using Eq. \ref{matrixelement} we have
	\be\label{rhsA}
	\mathcal{M}^{++}_{\uparrow \uparrow} + \mathcal{M}^{++}_{\downarrow \downarrow} \es 2\  (P^+)^2A_\G(q^2), \\
	\label{rhsB}
	\mathcal{M}^{++}_{\uparrow \downarrow} + \mathcal{M}^{++}_{\downarrow \uparrow} \es \frac{ i q^{(2)}}{m} \ (P^+)^2 B_\G(q^2) .
	\ee
 Using the two-particle LFWFs for a dressed quark state, we calculate the LHS of the above analytically; the details of the calculation of Eqs. \ref{rhsA} and \ref{rhsB} are given in Appendix \ref{appaAandB}. As stated in the introduction, the total $\overline{C}(q^2)$ should be zero when summed over quark and gluon due to the conservation of the EMT \cite{Lorce:2018egm}. However, non-zero contributions come both from the quark and gluon parts. Quark and gluon contributions to this GFF can be calculated as :
	\begin{align}
		\langle P^{\prime},S^{\prime}|	\partial_{\mu}\theta^{\mu\nu}_i(0)|P,S \rangle = iq^{\nu} \overline{C}_i(q^2)m\,\overline{U}(P^{\prime},S^{\prime})U(P,S).
		\label{cbar1}
	\end{align}

Taking $\nu=1$, we arrive at the following combination to extract $\overline{C}_\G(q^2)$,
	\begin{align}
		q_{\mu}\mathcal{M}^{\mu 1}_{\uparrow \downarrow} + q_{\mu}\mathcal{M}^{\mu 1}_{\downarrow \uparrow} = -i q^{(1)}q^{(2)}m\, \overline{C}_\G(q^2).
		\label{cbar3}   
	\end{align}
 
 The fourth GFF, $D_\G(q^2)=4C_\G(q^2)$ also known as the \emph{D}-term is obtained from the transverse component of the EMT. Both  $\overline{C}_G(q^2)$ as well as $D_\G(q^2)$ involve quark-gluon interaction in the operator structure after we eliminate the constrained fields. Details of this calculation are given in Appendix \ref{appaDandCbar}. We use the following linear combination of the transverse Lorentz indices to extract the \emph{D}-term,
	\be
	\label{rhsC}
	\mathcal{M}^{11}_{\uparrow \downarrow} +
	\mathcal{M}^{22}_{\uparrow \downarrow} +\mathcal{M}^{11}_{\downarrow \uparrow} + \mathcal{M}^{22}_{\downarrow \uparrow} \es
	i\bigg[B_\G(q^2)\frac{q^2}{4m}-D_\G(q^2)\frac{3q^2}{4m}+\overline{C}_\G(q^2) 2m\bigg]q^{(2)}.
	\ee 
	
	The final expression for the gluon GFFs are as follows: 
\be
 A_\G(q^2)\!\!\!\!\es\!\!\! \frac{g^2C_F}{8\pi^2}\left[\frac{29}{9}+\frac{4}{3}\ln\left(\frac{\Lambda^2}{m^2}\right)-\int dx\left(\left(1+\left(1-x\right)^2\right)+\frac{4m^2x^2}{q^2\left(1-x\right)}\right)\frac{\tilde{f_2}}{\tilde{f_1}}\right]\label{ag},\\
B_\G(q^2)\!\!\!\!\es\!\!\! -\frac{g^2C_F}{2\pi^2}\, \int dx\,\frac{m^2x^2}{q^2}\frac{ \tilde{f_2}}{ \tilde{f_1}}\label{bg},\\
D_\G(q^2) \!\!\!\!\es\!\!\! \frac{g^2C_F}{6\pi^2}\left[\frac{2 m^2}{3  q^2}+\int dx \frac{m^2}{ q^4 }\left(x\left(\left(2-x\right)q^2-4m^2x\right)\right)\right]\frac{\tilde{f_2}}{\tilde{f_1}}\label{dg},\\
\overline{C}_\G(q^2)\!\!\!\!\es\!\!\!   \frac{g^2C_F}{72\pi^2}\left[10 +9
\int dx~\left(x-\frac{4m^2x^2}{q^2\left(1-x\right)}\right)\frac{\tilde{f_2}}{\tilde{f_1}}-3\ln\left(\frac{\Lambda^2}{m^2}\right)\right]\label{cbarg},
\ee
where, 
\be
&&\tilde{f_1}:= \sqrt{1+\frac{4m^2x^2}{q^2\left(1-x\right)^2}}.\\
&&\tilde{f_2}:= ln\left(\frac{1+\tilde{f_1}}{-1+\tilde{f_1}}\right).
\ee

In order to calculate the analytical form of the GFFs we have used the integrals over the transverse  momentum as shown in Appendix \ref{appaint}.
It should be noted that a UV cut-off $\Lambda$ has been used to calculate the GFFs $A_\G(q^2)$ and $\overline{C}_\G(q^2)$ in the transverse momenta ($\bska$) integration, but for GFFs $B_\G(q^2)$ and $D_\G(q^2)$, the $\bska$-integration is found to be convergent, and no cutoff is used. This cutoff introduces a renormalization scale dependence on the GFFs  $A_\G(q^2)$ and $\overline{C}_\G(q^2)$; in our approach, this has its origin in the transverse momenta ($\bska$) integration. But the total GFFs of a system are independent of such scale dependence \cite{Polyakov:2018zvc}. Indeed our total GFFs are independent of the UV cut-off since any $\Lambda$ dependent contribution from the gluon exactly cancels with the corresponding quark contribution. The calculation of quark contribution to GFFs has been explored in our previous work \cite{More:2021stk}.
		\section{Numerical analysis: Gravitational form factors}
  
		 In this section, we show the plots of the GFFs  $A_\G(q^2)$, $B_\G(q^2)$, $D_\G(q^2)$ and $\overline{C}_\G(q^2)$ listed in Eqs. \ref{ag} - \ref{cbarg} as a function of the momentum transferred squared $(q^2)$. We take the mass of the dressed quark to be  $m = 0.3 ~\mathrm{GeV}$ and $g= C_F = 1$. We use the cutoff $\Lambda = 2~\mathrm{GeV}$ for the calculation of both quark and gluon GFFs.
		
\begin{figure}[ht]
			\begin{minipage}{0.45\linewidth}
				\includegraphics[scale=0.4]{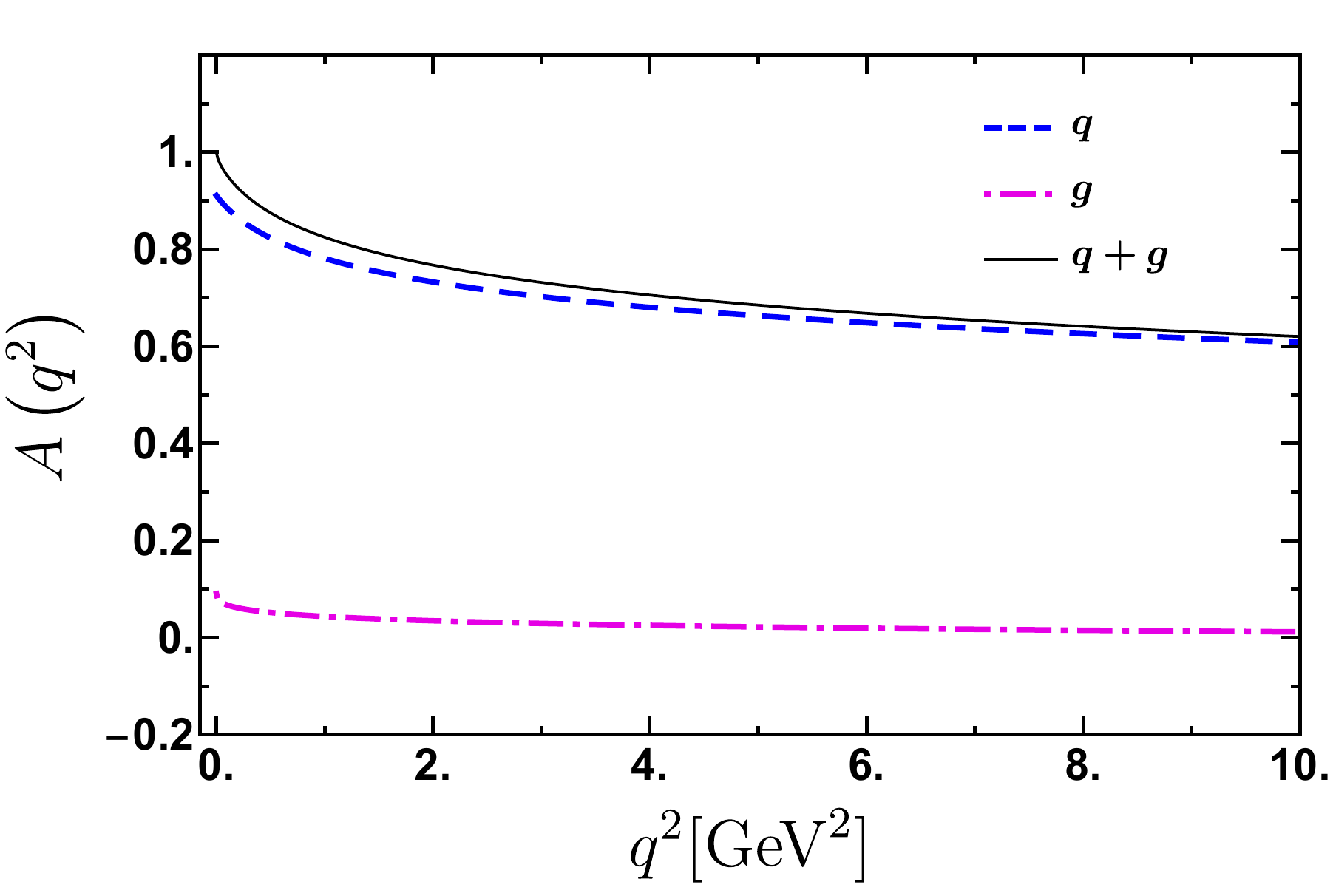}
			\end{minipage}
			\begin{minipage}{0.45\linewidth}
				\includegraphics[scale=0.4]{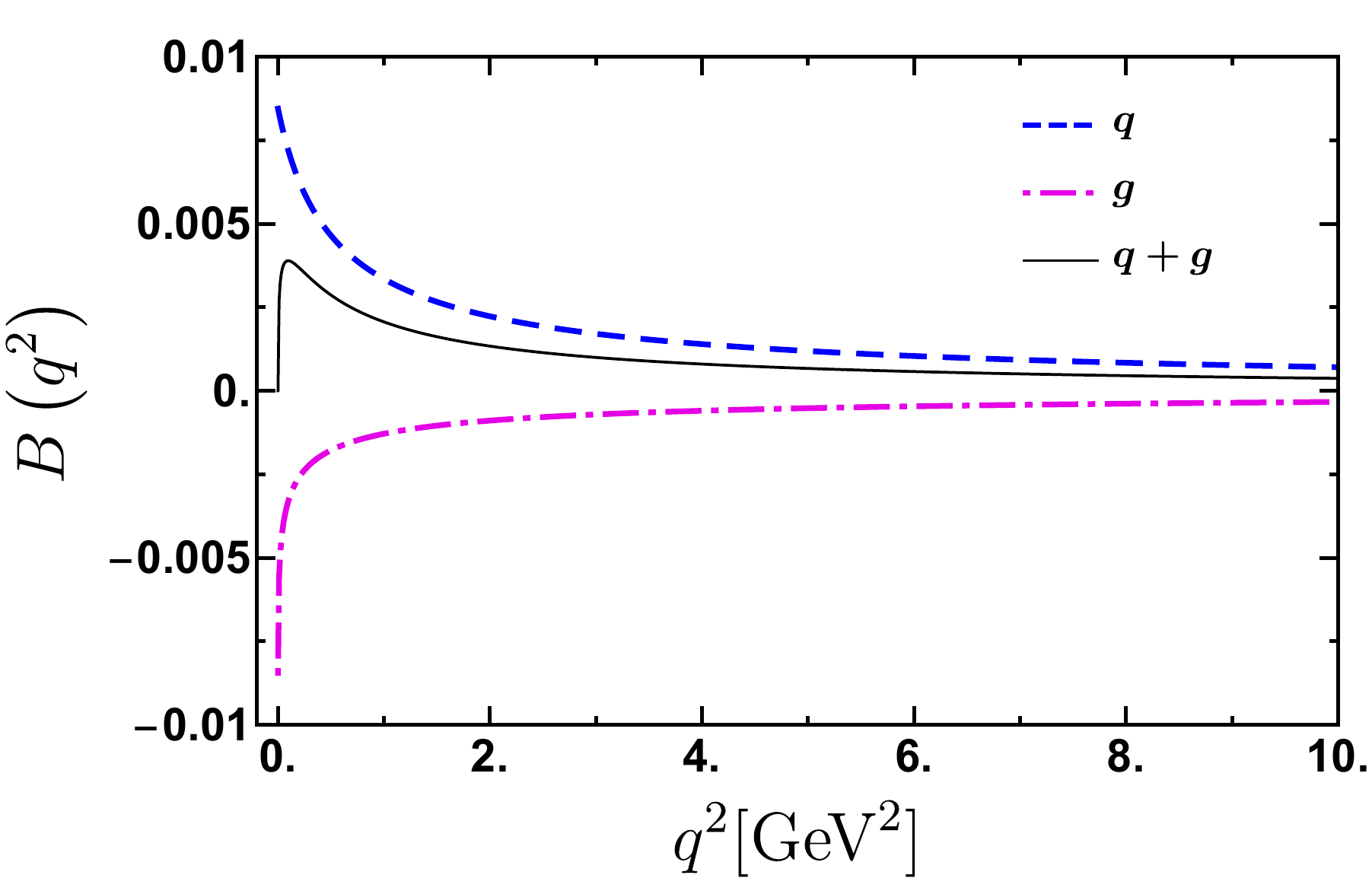}
			\end{minipage}  
			\caption{Plots of the GFF $A(q^2)$ and $B(q^2)$ as a function of $q^2$. The dashed blue curve and the dot-dashed magenta curve are for the quark ($q$) and gluon ($g$) form factors respectively. The solid black curve is for the sum of quark and gluon ($q+g$) contribution. Here $m = 0.3 ~\mathrm{GeV}$ and $\Lambda = 2~\mathrm{GeV}$.}
			\label{totalfiggffAnB}
		\end{figure}		
		In Fig. \ref{totalfiggffAnB}, we plot the total GFFs $A(q^2)$ and $B(q^2)$ as a function of $q^2$ along with the individual quark and gluon contributions. As discussed above, $A(q^2)$ for the quark and gluon depend on $\Lambda$, although the total $A(q^2)$ is independent of this cutoff. These form factors have been calculated for a 
 dressed electron system in QED as well as in Yukawa theory in \cite{Brodsky:2000ii}. The QED limit can be obtained from our calculation by taking suitable values of the parameters. Our results for $A(q^2)$ and $B(q^2)$ agree  with this reference. Our result for the GFF $A_\G(q^2)$ matches with an existing result which was calculated through GPDs in the massless limit \cite{Chakrabarti:2004ci}. Lattice study of the gluon contribution to GFFs of the nucleon, $A_\G(q^2)$ and $B_\G(q^2)$ have been made in \cite{Deka:2013zha,Shanahan_2019}. Our results agree qualitatively for $A_G(q^2)$, on the lattice the gluon contribution to the GFF $B_\G(q^2)$ in \cite{Shanahan_2019} is found to be slightly positive, although in \cite{Deka:2013zha} it is shown to be negative. As seen in the plot above, in the dressed quark state, the contribution from the gluon to the GFF $B(q^2)$ is negative. As a consequence of Poincare invariance the total GFFs $A(q^2)$ and $B(q^2)$ satisfy the sum rules $A(0)=1$ and $B(0)=0$ respectively \cite{ Lorce:2015lna,Lowdon_2017}. These sum rules are satisfied which can be seen in fig. \ref{totalfiggffAnB}. This can be obtained analytically also and is explicitly shown in Appendix \ref{appaAandB}. 
 The condition on GFF $A(q^2)$ physically gives the momentum sum rule, and  for $B(q^2)$ it means that the anomalous gravitomagnetic moment vanishes for a spin-$1/2$ system \cite{Polyakov:2018zvc}. Together these two GFFs satisfy Ji's sum rule  \cite{PhysRevLett.78.610}:
  \beq
   J_{Q,G}(0)= \frac1{2} (A_{Q,G}(0)+B_{Q,G}(0)).
  \eeq
  
 The above relation gives the contribution to the total angular momentum from the quark/gluon.

   \begin{figure}[ht]
			\begin{minipage}{0.45\linewidth}
				\includegraphics[scale=0.4]{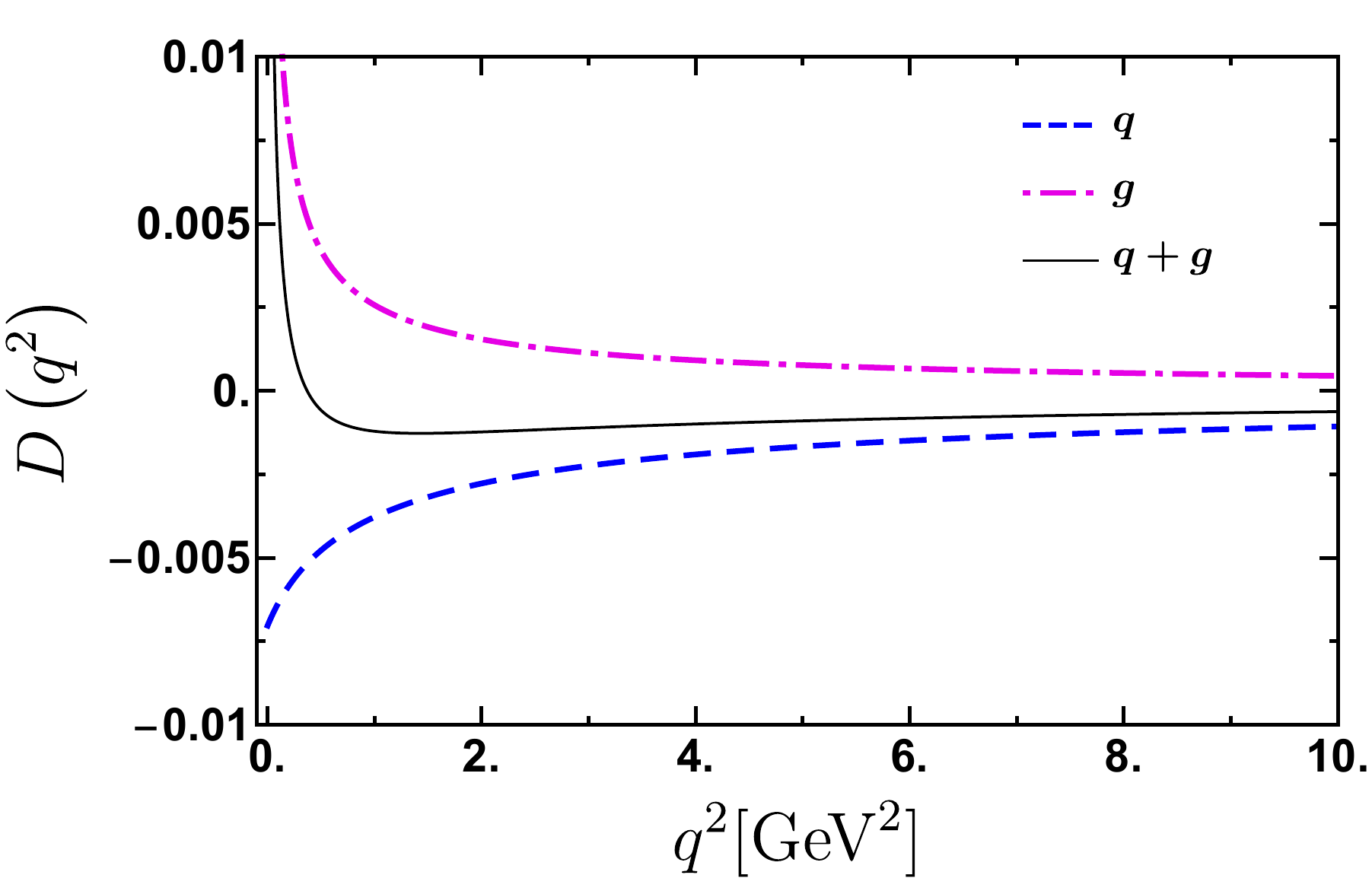}
			\end{minipage}
			\begin{minipage}{0.45\linewidth}
				\includegraphics[scale=0.4]{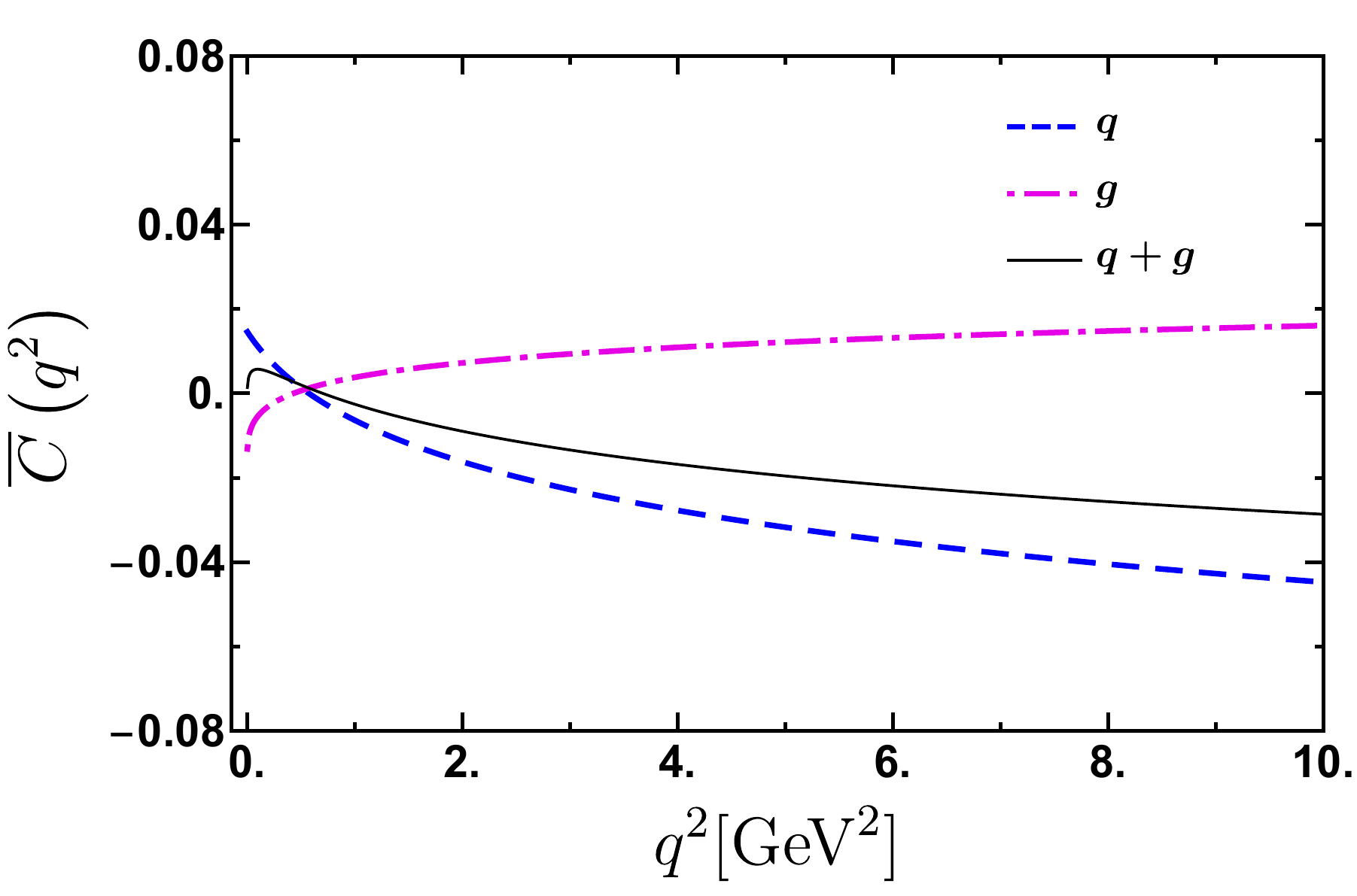}
			\end{minipage}  
			\caption{Plots of the GFF $D(q^2)$ and $\overline{C}(q^2)$ as a function of $q^2$. The dashed blue curve and the dot-dashed magenta curve are for the quark ($q$) and gluon ($g$) form factors respectively. The solid black curve is for the sum of quark and gluon ($q+g$) contribution. Here $m = 0.3 ~\mathrm{GeV}$ and $\Lambda = 2~\mathrm{GeV}$.}
			\label{totalfiggffDnCbar}
		\end{figure}

\begin{figure}[htbp]
\begin{minipage}{0.45\linewidth}
\includegraphics[scale=0.35]{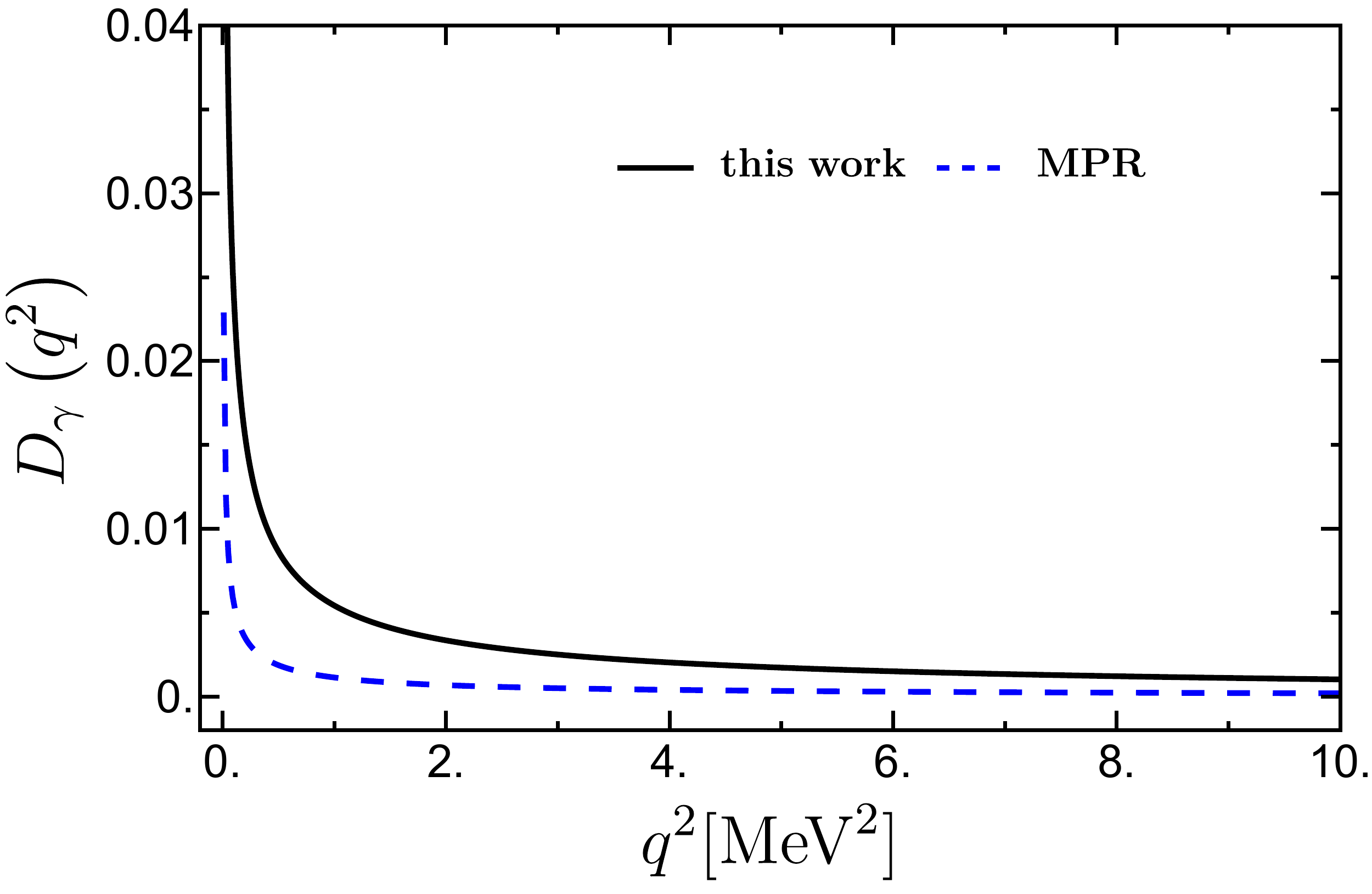}
\end{minipage}
\caption{Plot of photon GFF $ D_{\gamma}(q^2)$ as function of $q^2$. The blue dashed curve is the result for the photon \emph{D}-term as shown in Ref.~\cite{Metz:2021lqv} (MPR) and the solid black line is our \emph{D}-term with the same parameter. Here we set $m=0.511$ MeV, $\alpha=\frac{1}{137}$. }
\label{Dgammacompare}
\end{figure}

  In Fig. \ref{totalfiggffDnCbar}, we plot the GFFs $D(q^2)$ and $\overline{C}(q^2)$ as  functions of $q^2$ for the contribution of the quark, the contribution of the gluon and the total contribution of quark and gluon parts of the EMT. In this case the GFF $\overline{C}_{\G}(q^2)$ is found to change from negative to positive value as $ q^2$ increases and the value of 
		$\overline{C}_{\G}(0)=-0.0146$. The quark counterpart shows exactly opposite behaviour and the value of $\overline{C}_{\Q}(0)=0.0146$. The individual GFFs $\overline{C}_i(q^2)$ depends on the UV cut-off $\Lambda$. However the total $\overline{C}(q^2)$ is independent of $\Lambda$, as expected. We observe that the total $\overline{C}(q^2)$ is non-zero, except at $q^2=0$, $\overline{C}(0)=0$ as shown in Eq.~\ref{cbarsum}. The total $\overline{C}(q^2)$ however, is expected to be zero due to the conservation of EMT. $\overline{C}(q^2)$ depends on the transverse component of the EMT. As discussed in the Appendix \ref{appaDandCbar}, in the calculation of the matrix elements for a dressed quark state in two-component formalism, we have not included the contribution from the terms that have $k^+=0$ for either the quark or the gluon. It is possible that due to this, $\overline{C}(q^2)$ is non-zero in our calculation. A careful inclusion of 
  the light-front zero modes is required to see the result coming from the conservation of the full EMT, this is beyond the scope of the present work. The slope of the $\overline{C}_\Q$ is steeper than $\overline{C}_{\G}$. As discussed in the introduction, the \emph{D}-term is not related to any Poincare generators and therefore is not constrained. It is related to the pressure and shear force inside the nucleon. The value of the $D_{\G}(q^2)$ is found to be positive and divergent as $q^2$ approaches zero.		 The total $D(q^2)$ is divergent as $q^2 \rightarrow 0$ as well. The total $D(q^2)$ is found to be negative in the chosen range as expected for a bound state except in the region close to $q^2=0$, where the \emph{D}-term is positive. The GFFs for an electron in QED have been calculated in a Feynman diagram approach in \cite{Freese:2022jlu, Metz:2021lqv}. The \emph{D}-term for the photon part of the EMT has been found to be divergent. This divergence was related to the infinite range of the Coulomb interaction \cite{Berends:1975ah,Milton:1976jr}.  In our one loop calculation in QCD we also see a divergent behaviour. The \emph{D}-term for the photon can be obtained from our result by setting the QED values for the parameters.  We observe that our result for the \emph{D}-term for the photon shows qualitatively similar behaviour with \cite{Metz:2021lqv}. The comparison has been shown in Fig. \ref{Dgammacompare}. The quantitative difference may be due to the fact that the light-front zero modes are not included in our calculation. This requires further investigation, possibly in another publication.  
  
\section{Pressure and Shear Force distributions}

The \emph{D}-term encodes information about the pressure and shear distributions inside the nucleon \cite{Polyakov:2018zvc}. The pressure and shear force distributions are encoded in the transverse component of the EMT through the pure-stress tensor~\cite{Freese:2021czn}. The gluon contribution to the \emph{D}-term is vital for a complete understanding of the quark-gluon dynamics associated with the transverse components of the EMT.
As discussed in the introduction, such distributions in the literature have been defined in different frames. The spatial distributions in the Breit frame do not have an interpretation of spatial density \cite{Freese:2021czn}. There is in general an ambiguity to define a three-dimensional spatial distribution of a local property of the nucleon which has a size much less than the Compton wavelength \cite{Jaffe:2020ebz}; this ambiguity comes while localizing the nucleon to define such distributions. In light-front formalism, we can define 2D distributions at constant light front time $x^+=0$ by taking a two-dimensional Fourier transform. In the Drell-Yan frame when the momentum transfer is purely in the transverse direction, such distributions have density interpretation \cite{Freese:2021czn}.
  In this work, we calculate the distributions in the form of Fourier transform of $\bsq$ to the impact parameter space $\bsb$ in the Drell-Yan frame following the approach of \cite{Lorce:2018egm}. In references \cite{Burkardt:2002hr,Burkardt:2000za} boost invariant impact parameter distributions (IPDFs) of quarks and gluons has been introduced by taking 2D Fourier transform of the GPDs, using the fact that, in light-front framework, the transverse boosts are Galilean in nature. The expressions for pressure and shear distributions in two dimensions \cite{Freese:2021czn} are 
\be
\label{Prefun}
p_i(b^{\perp}) \es \frac{1}{8m b^{\perp}} \frac{d}{db^{\perp}} \left[b^{\perp} \ \frac{d}{db^{\perp}} D_{i}(b^{\perp})\right]-m\ \overline{C}_i(b^{\perp}),\\
s_i(b^{\perp})\es-\frac{b^{\perp}}{4m} \frac{d}{db^{\perp }}\left[ \frac{1}{b^{\perp }} \frac{d}{db^{\perp }} D_i(b^{\perp})\right],
\label{Prefun2}
\ee

where
\be
\label{fgff}
F(b^{\perp})
\es  \frac{ 1}{(2\pi)^2}~\int d^2 \bsq \ e^{-i\bsq \bsb} \mathcal{F}(q^2) \nn \\
&=&\frac{1}{2\pi}\int_0^{\infty} d  q^{\perp} ~ q^{\perp} J_0\left( q^{\perp} b^{\perp}\right)\mathcal{F}(q^2),
\ee
		
where $\mathcal{F}=\left(A_i,B_i,D_i, \overline{C}_i\right)$, $i\equiv(Q, G)$. $J_0$ is Bessel's function of zeroth order. $m$ is the mass of the dressed quark state. The last term in Eq.~\ref{Prefun} will not be there for the pressure distribution derived from conserved total EMT.
In order to avoid infinities at intermediate steps, we use a wave packet state to calculate these spatial distributions \cite{Diehl:2002he,Chakrabarti:2005zm}.   The dressed quark state confined in transverse momentum space with definite longitudinal momentum can be written as  
		\be
		\frac{1}{16\pi^3}\int \frac{d^2 \bsp dp^+}{p^+}\phi\left( p\right) \mid p^+,\bsp,\lambda \rangle ,
		\ee
		with $\phi(p)=p^+\ \delta(p^+-p_0^+)\ \phi\left(p^{\perp}\right)$.
		We use a Gaussian wave packet in transverse momenta  at fixed longitudinal light-front momentum. 
		\be\label{gaussian}
		\phi\left(p^{\perp}\right)
		= e^{-\frac{p^{\perp2}}{2\Delta^2}},
		\ee 
		where $\Delta$ is the width of Gaussian.
		\begin{figure}[ht]
			\begin{minipage}{0.45\linewidth}
				\includegraphics[scale=0.4]{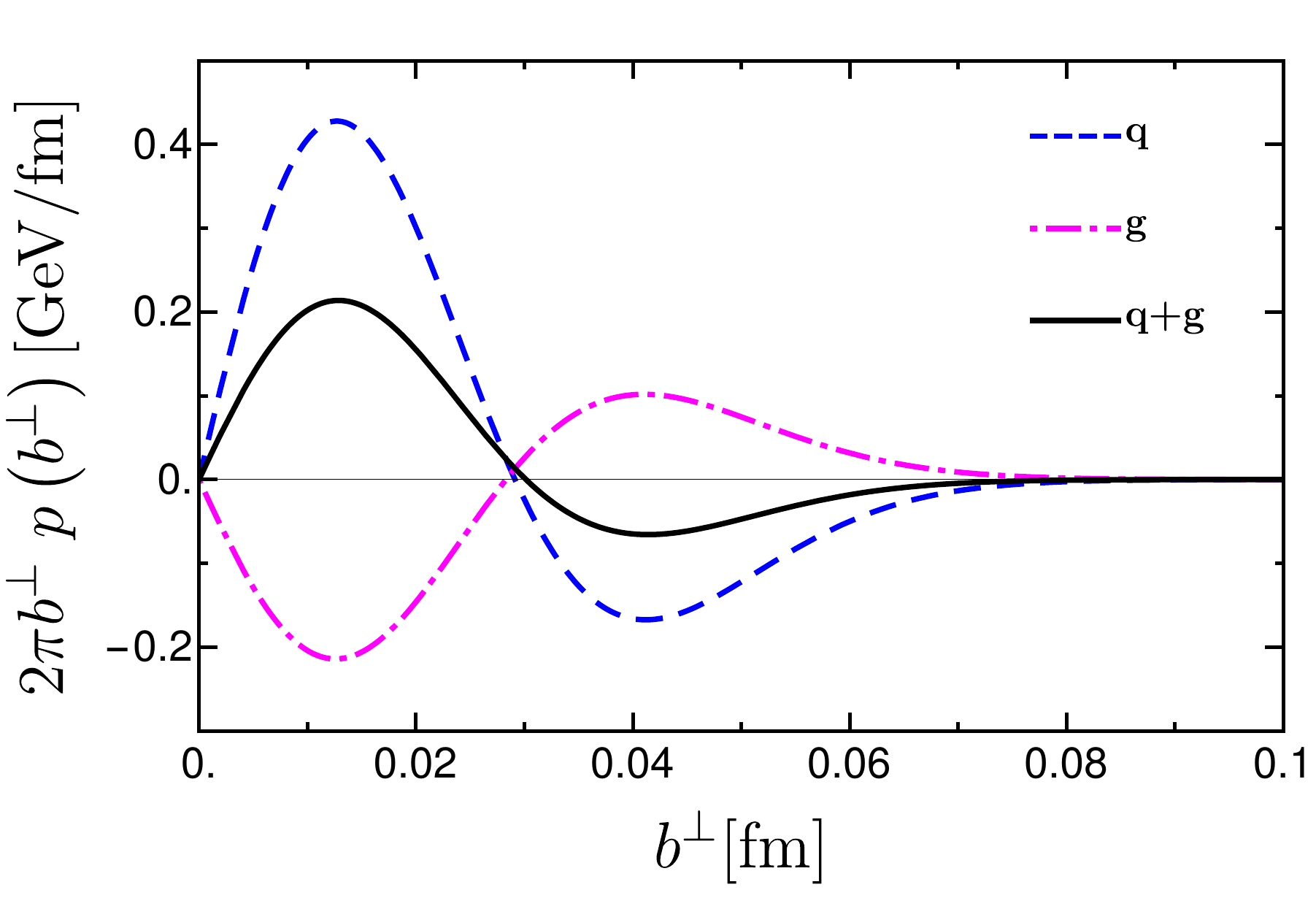}
			\end{minipage}
			\begin{minipage}{0.45\linewidth}
				\includegraphics[scale=0.4]{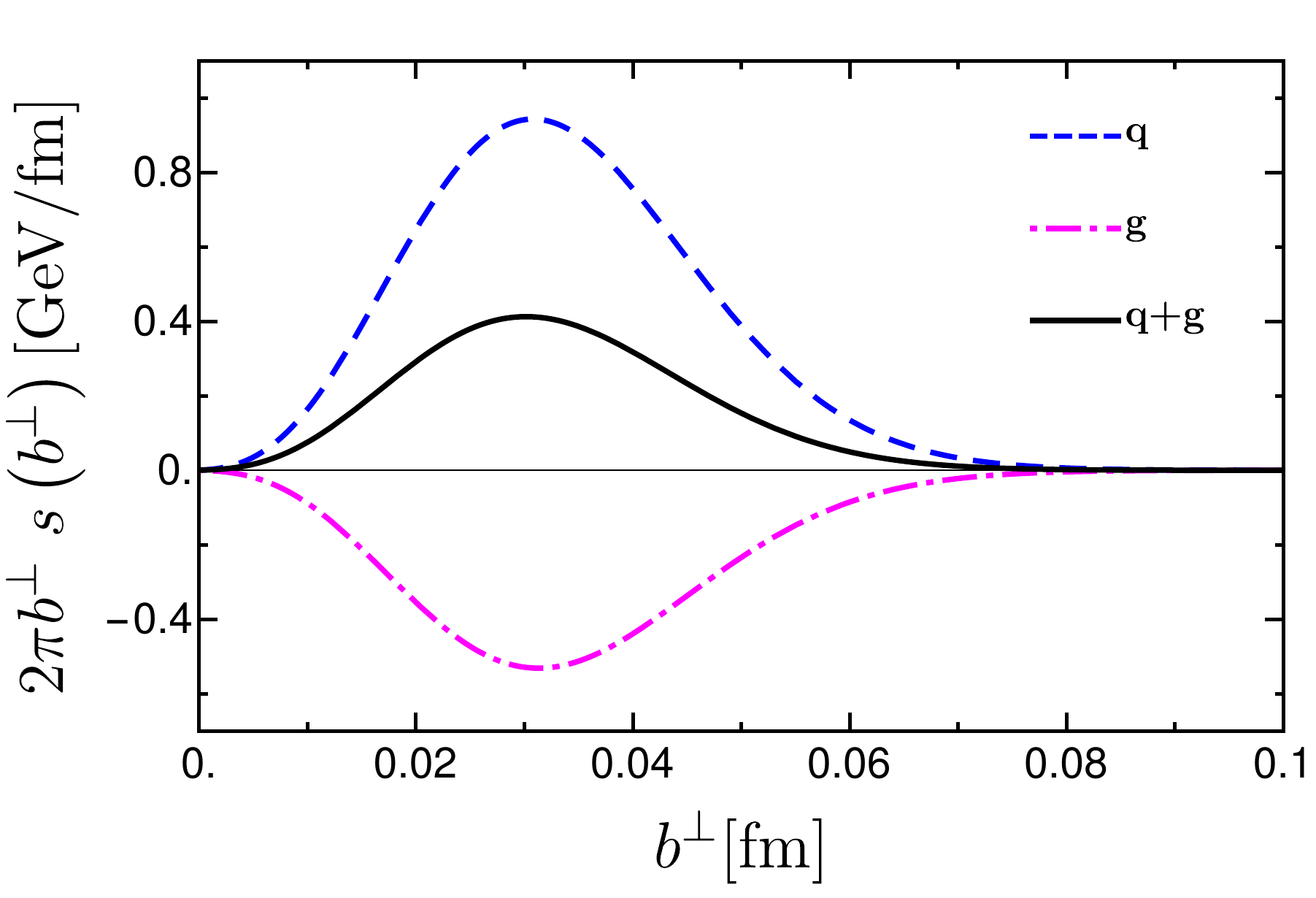}
			\end{minipage}  
			\caption{Plots of the pressure distribution $2\pi b^{\perp}~p(b^{\perp})$ and the shear force distribution $2\pi b^{\perp}~s(b^{\perp})$ as a function of $b^\perp$. The dashed blue curve and the dot-dashed magenta curve are for the quark ($q$) and gluon ($g$) contributions respectively. The solid black curve is for the sum of quark and gluon ($q+g$) contribution. Here $\Delta=0.2$.}
			\label{pns}
		\end{figure}
	
		  This state  provides Fourier-transformed pressure and shear distribution and also smooth plots. The combination of pressure and shear gives the normal and the tangential forces experienced by a ring of radius $b^\perp$  :
		\be
		\label{Fn-Ft}
		F_n(b^{\perp})\es2\pi b^{\perp}\left( p(b^{\perp}) + \frac{1}{2} s(b^{\perp})\right),
		\\
		F_t(b^{\perp})\es 2\pi b^{\perp} \left( p(b^{\perp}) - \frac{1}{2} s(b^{\perp})\right).
		\ee

   Here we have suppressed the index $i$ for quark/gluon. In Fig. \ref{pns} we have shown  the plots of $2 \pi b^{\perp} ~p(b^{\perp})$ and $2 \pi b^{\perp}  ~s(b^{\perp})$.  For these plots, we have taken the Gaussian width $\Delta=0.2$. We have shown three plots for quark, gluon and the total contribution. 
 The quark and gluon contributions to the pressure distribution have one node each that coincides. The quark contribution is larger than the gluon and hence the total contribution mimics the behaviour shown by the quark contribution. The profile of the total pressure curve indicates that at the centre of the two-particle system, there is a positive core, and there is negative pressure distribution towards the boundary. This behaviour of the pressure distribution is essential for a stable system, which is a repulsive core balanced by the confining pressure in the outer region \cite{Burkert:2018bqq}. This behaviour of the pressure distribution is similar to what has been observed for a nucleon for example from fits the JLab data  \cite{Burkert:2018bqq}. The total pressure profile of the dressed quark satisfies the Von Laue stability condition  \cite{Freese:2021czn} for which at least a single node should be present in the pressure distribution (left panel of Fig. \ref{pns}).
		\be\int d^2 \bsb ~p(\bsb) = 0.
		\ee
 	\begin{figure}[ht]
 	\begin{minipage}{0.45\linewidth}
 		\includegraphics[scale=0.4]{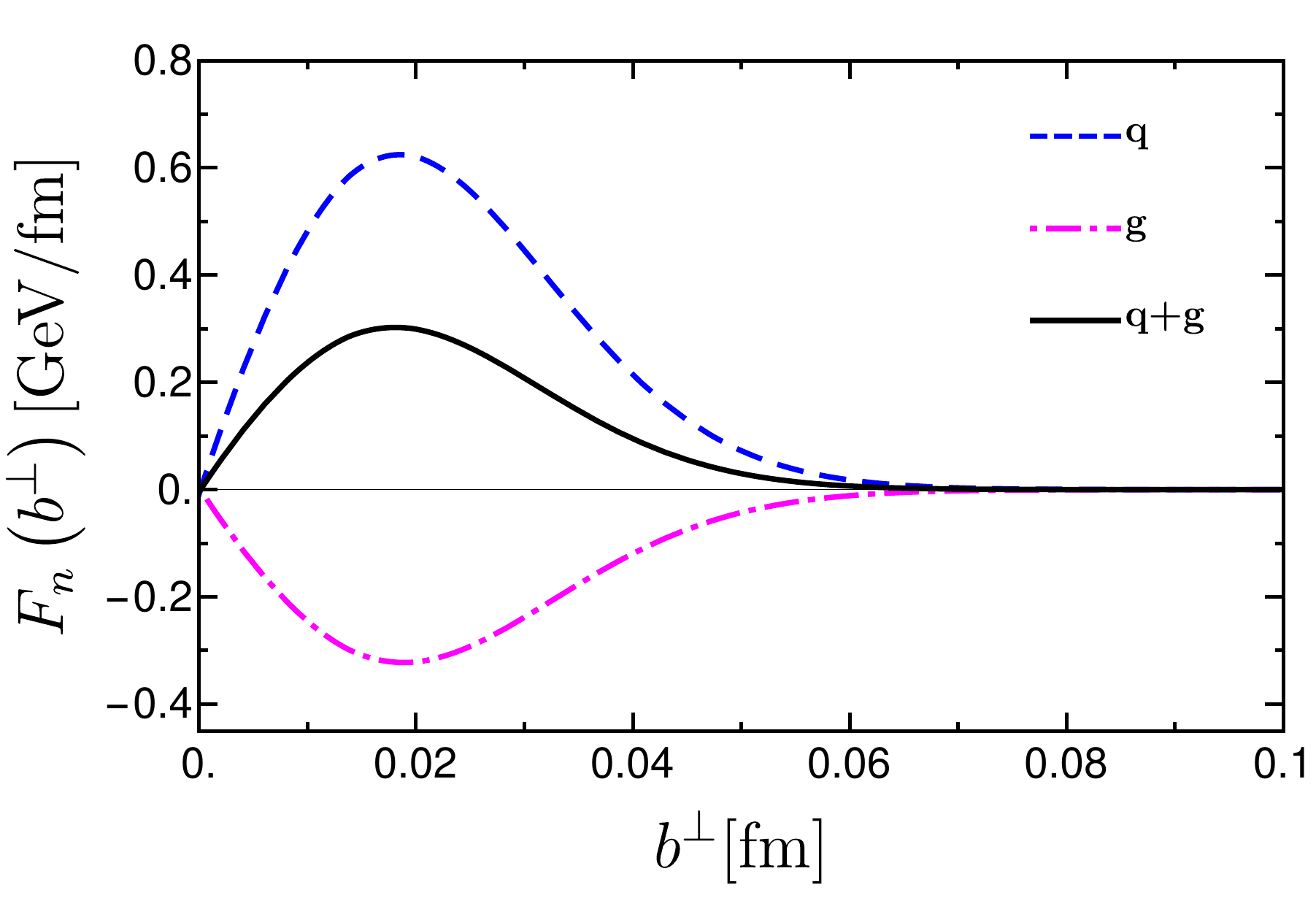}
 	\end{minipage}
 	\begin{minipage}{0.45\linewidth}
 		\includegraphics[scale=0.4]{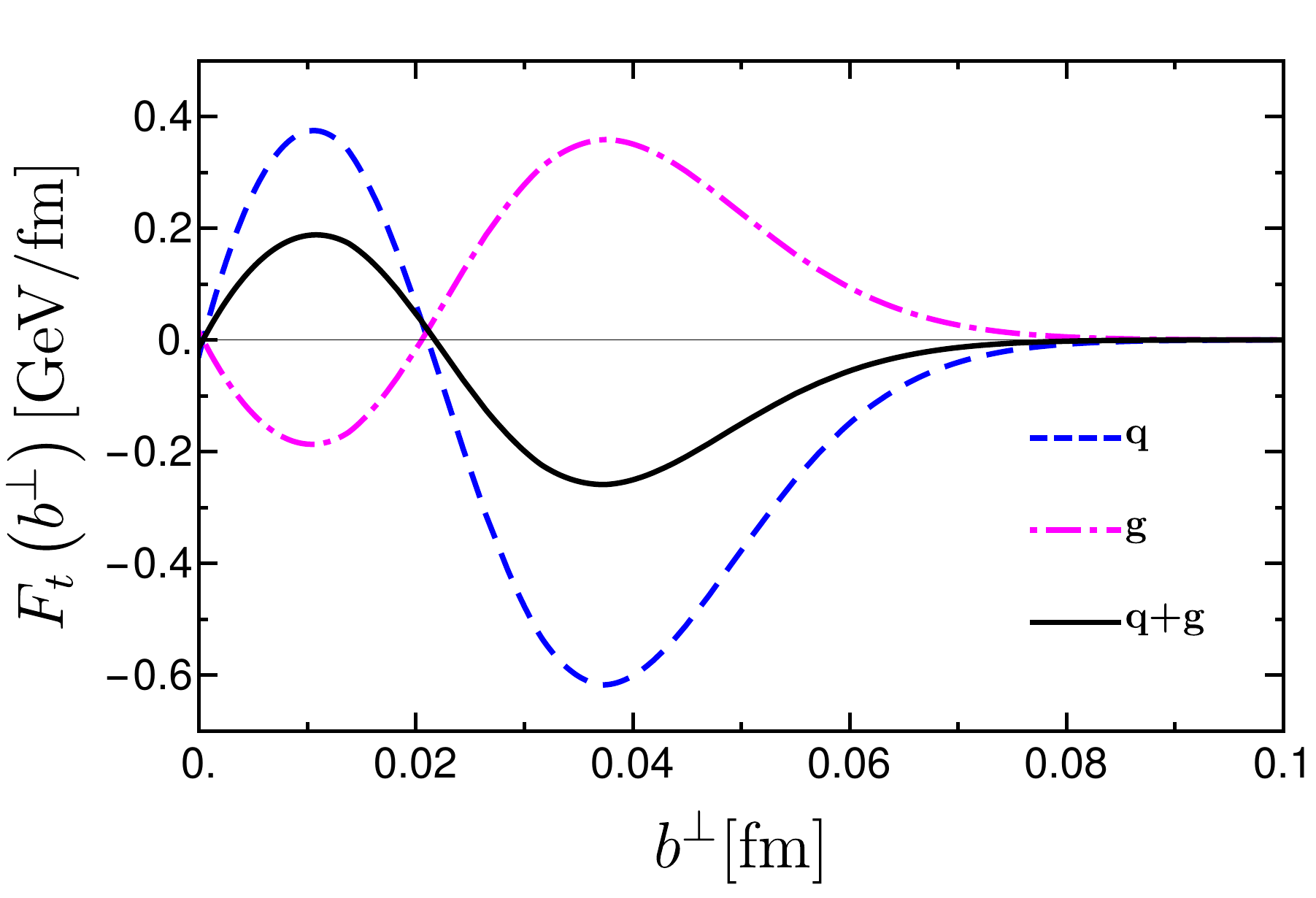}
 	\end{minipage}  
 	\caption{Plots of the normal force  $F_n(b^{\perp})$, and the tangential force $F_t(b^{\perp})$ as a function of $b^{\perp}$. The dashed blue curve and the dot-dashed magenta curve are for the quark ($q$) and gluon ($g$) contributions respectively. The solid black curve is for the sum of quark and gluon ($q+g$) contributions. Here $\Delta=0.2$.}
 	\label{fnft}
 \end{figure}

The shear force due to gluon turns out to be negative, unlike the quark shear force which is a positive definite quantity. The total shear force resembles that of a stable hydrostatic system \cite{Polyakov:2018zvc}(right panel of Fig. \ref{pns}).

		Fig. \ref{fnft} shows the normal force $F_n$ and tangential force $F_t$ in the impact parameter space. The gluon part of normal force $F_n$ is found to be negative, but the total normal force is positive. This means that the system is stable under collapse \cite{Polyakov:2018zvc}. The nature of the gluon contribution to the tangential force $F_t$ is similar to the quark but opposite in sign. The total tangential force is zero at the center of the two-particle system; positive near the core and negative towards outer region,  which keeps stability in the tangential direction.
		
\section{Energy density and pressure distributions}
		
The two-dimensional Galilean energy density, radial pressure, tangential pressure, isotropic pressure, and pressure anisotropy, are defined in Ref \cite{Lorce:2018egm}
\be \label{Genergy}
\mu_i(b^{\perp}) \es m \left [\frac{1}{2} A_i(b^{\perp})+ \overline{C}_i(b^{\perp}) + \frac{1}{4m^2}\frac{1}{b^{\perp}}\frac{d}{db^{\perp}} \left( b^{\perp}\frac{d}{db^{\perp}}\left[ \frac{1}{2}B_i(b^{\perp}) - 4C_i(b^{\perp}) \right]\right) \right], \\
\label{radialP}
\sigma^r_i(b^{\perp}) \es m  \left[
-\overline{C}_i(b^{\perp}) + \frac{1}{m^2} \frac{1}{b^{\perp}} \frac{d C_i(b^{\perp})}{db^{\perp}} 
\right],  \\
\label{tangentialP}
\sigma^t_i(b^{\perp}) 
\es m  \left[-\overline{C}_i(b^{\perp}) + \frac{1}{m^2} \frac{d^2C_i(b^{\perp})}{d b^{\perp2}} 
\right] \label{mp}, \\ 
\label{totalP}
\sigma_{i}(b^{\perp}) \es m  \left[
-\overline{C}_i(b^{\perp}) + \frac{1}{2}\frac{1}{m^2} \frac{1}{b^{\perp}} \frac{d}{db^{\perp}}\left(b^{\perp} \frac{d \ C_i(b^{\perp})}{d b^{\perp}} \right)
\right] , \\
\label{shearlike}
\Pi_i(b^{\perp})\es  m  \left[ -\frac{1}{m^2} b^{\perp} \frac{d}{db^{\perp}}\left(\frac{1}{b^{\perp}} \frac{dC_i(b^{\perp})}{d b^{\perp}} \right)\right]
\ee
 Eqs. (\ref{totalP}) and (\ref{shearlike}) agree with  Eqs. (\ref{Prefun}) and  (\ref{Prefun2}) above. 
The isotropic pressure and the pressure anisotropy can be defined in terms of radial  pressure and tangential pressure as follows
		
\begin{align}
\sigma_i=&~ \frac{(\sigma^r_i+\sigma^t_i)}{2},\\
\Pi_i=&~ \sigma^r_i-\sigma^t_i.
\label{sigmapi}
\end{align}

The distributions defined in Eq.~\ref{Genergy}-\ref{shearlike} are studied in the impact parameter space and hence we take the Fourier transform as shown in Eq.~\ref{fgff}. As before we choose a Gaussian wave packet state  with a width of $\Delta = 0.2$.

\begin{figure}[ht]
	\includegraphics[scale=0.4]{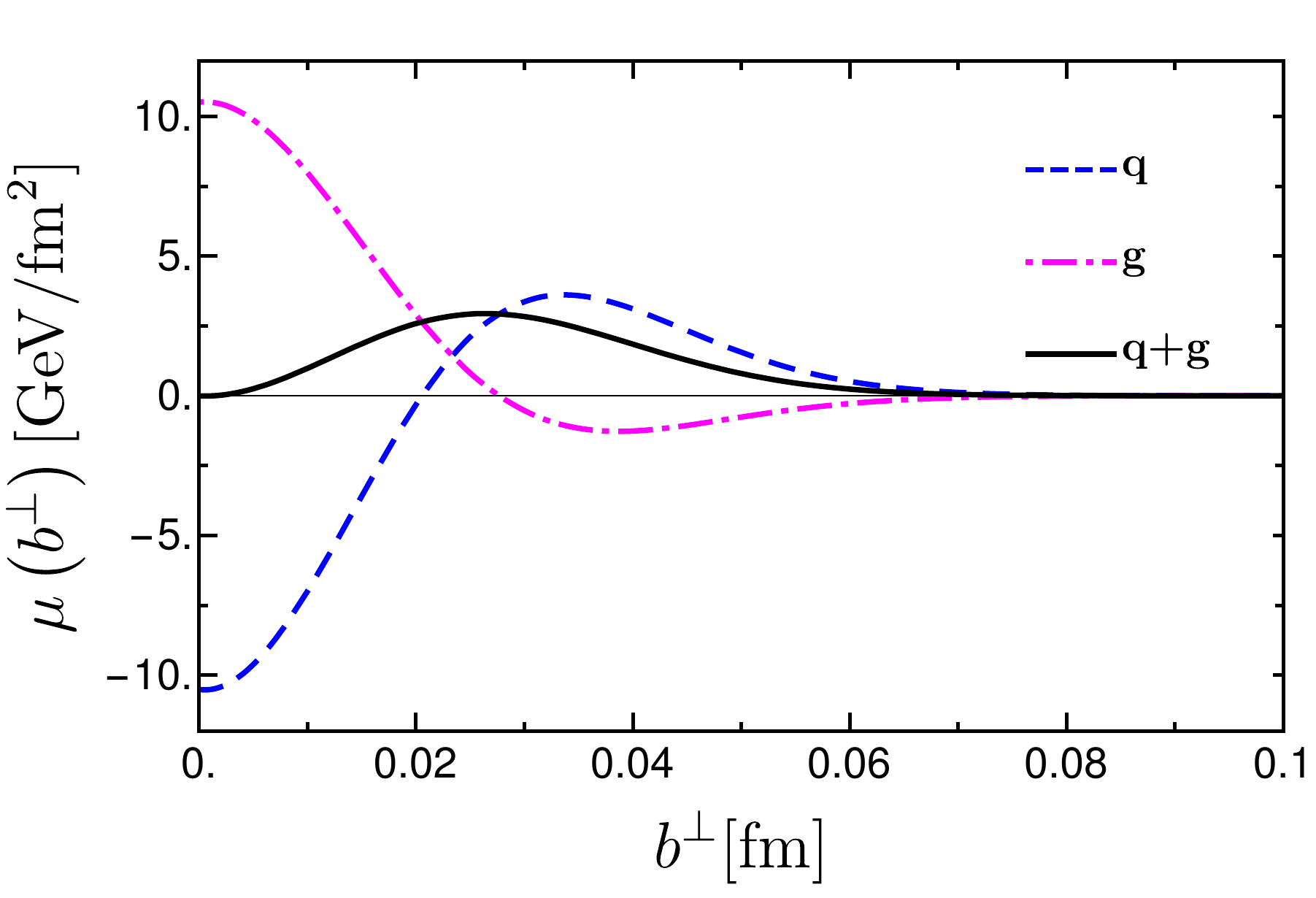} 
	\caption{Plot of the energy density $\mu(b^{\perp})$ as a function of $b^{\perp}$. The dashed  blue curve and the dot-dashed magenta curve are for the quark ($q$) and gluon ($g$) contributions respectively. The solid black curve is for the sum of quark and gluon ($q+g$) contribution. Here $\Delta=0.2$.}
	\label{mu}
\end{figure}
		
We show our result for the energy density in the impact parameter space in Fig.~\ref{mu}. The Galilean energy density involves all four GFFs as seen from Eq.~\ref{Genergy}. The energy density for quark shows a negative region near the center  whereas the gluon energy density has a negative region towards the peripheral region away from the center. The energy density of the gluon is positive near the center and as we move away from the center the quark contribution becomes positive. However, the total energy density remains a positive definite quantity over the entire range. The blue dashed curve and the dotted magenta curve intersect at around $b^\perp  \approx 0.025~\mathrm{fm}$ which is the point where the quark and gluon energy density share the exact same non-zero value. The total energy density has a peak centered around $b^\perp  \approx 0.03~\mathrm{fm}$, here $b^\perp $ indicates the magnitude of $\bsb$.

\begin{figure}[ht]
\begin{minipage}{0.45\linewidth}
\includegraphics[scale=0.4]{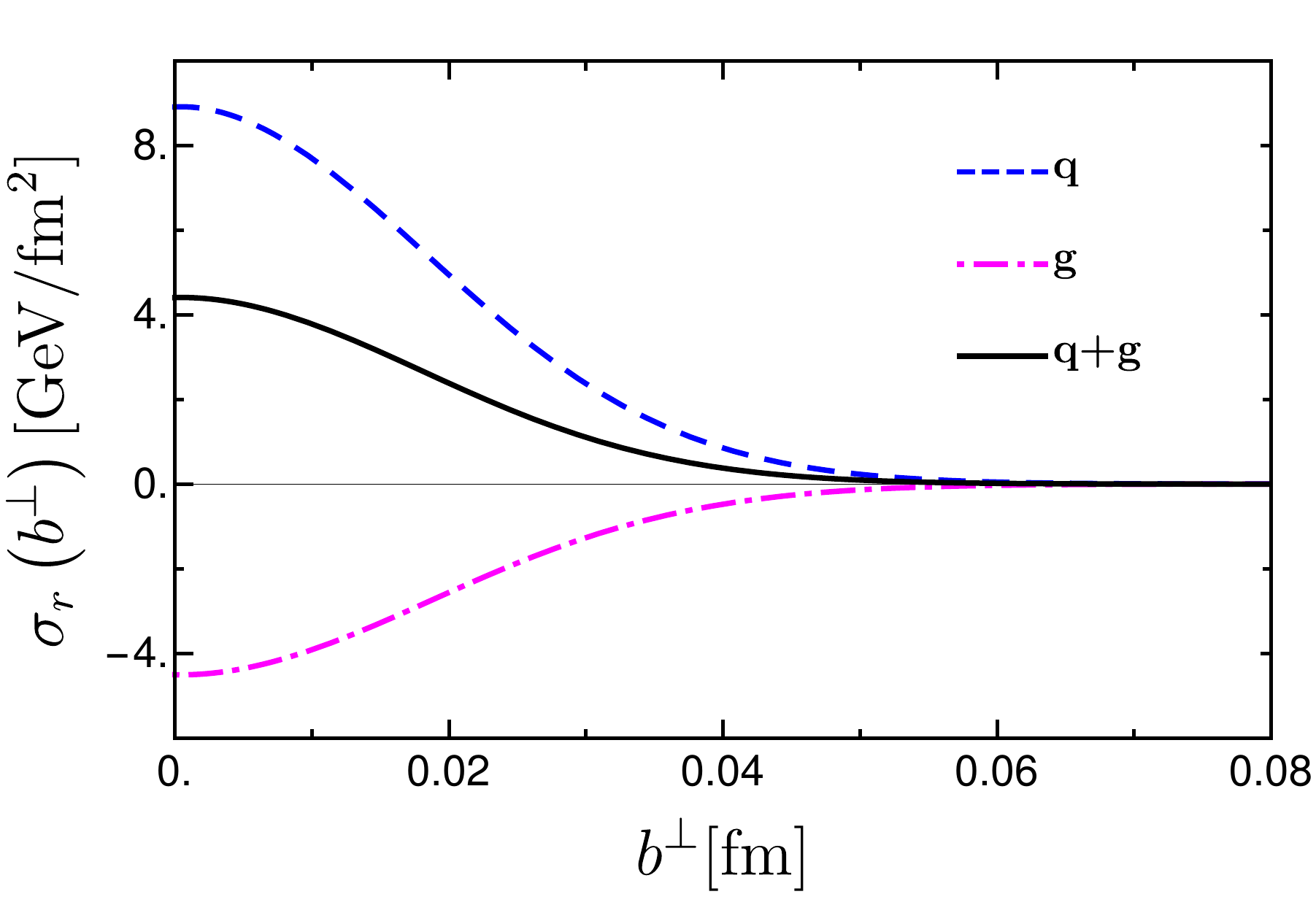}
\end{minipage}
\begin{minipage}{0.45\linewidth}				\includegraphics[scale=0.4]{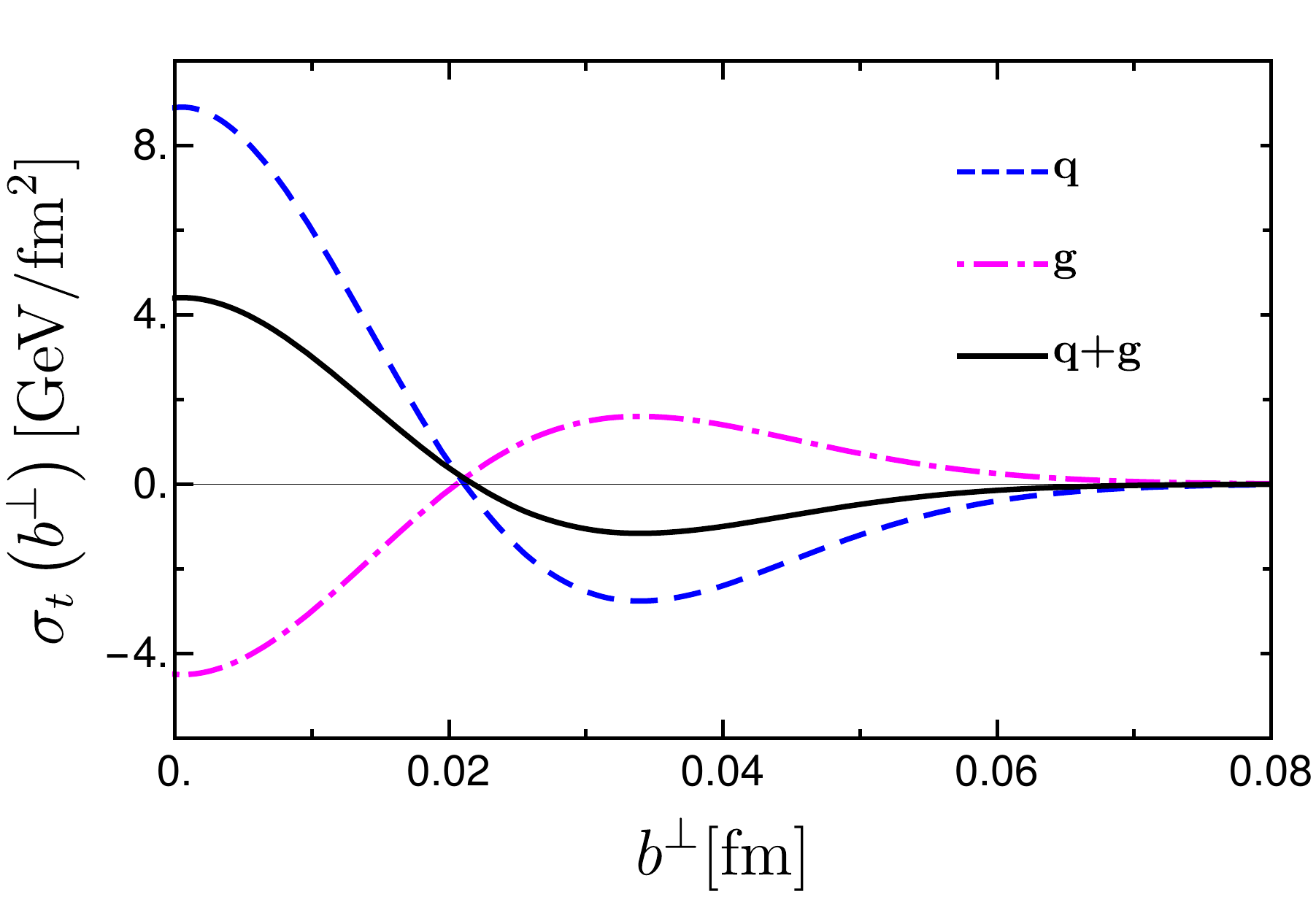}
\end{minipage}  
\caption{Plots of the 2D radial pressure $\sigma_r(b^{\perp})$ and the tangential pressure $\sigma_t(b^{\perp})$ as a function of $b^{\perp}$. The dashed blue curve and the dot-dashed magenta curve are for the quark ($q$) and gluon ($g$) contributions respectively. The solid black curve is for the sum of quark and gluon ($q+g$) contribution. Here $\Delta=0.2$.}
\label{sigmaRsigmaT}
\end{figure}
		
  Fig.~\ref{sigmaRsigmaT} shows the plot for the radial and tangential pressure as defined in Eq.~\ref{radialP} and Eq.~\ref{tangentialP} respectively. The quark radial pressure has a maximum at the center of the impact parameter space and it falls off gradually becoming zero around $b^{\perp} \approx 0.06~\mathrm{fm}$. On the other hand, the gluon radial pressure exhibits a negative value at the center. The quark contribution dominates over the gluon such that the total radial pressure stays positive.
  The tangential pressure for both quark and gluon show positive as well as negative regions. The tangential pressure due to gluon is found to be negative till $b^{\perp} \approx 0.02 ~\mathrm{fm}$ and positive after that. The quark behavior is inverted compared to the gluon case. However, the total tangential pressure mimics the quark behavior which indicates that the quark contribution is larger compared to the gluon.
  
\begin{figure}[ht]
\begin{minipage}{0.45\linewidth}
\includegraphics[scale=0.4]{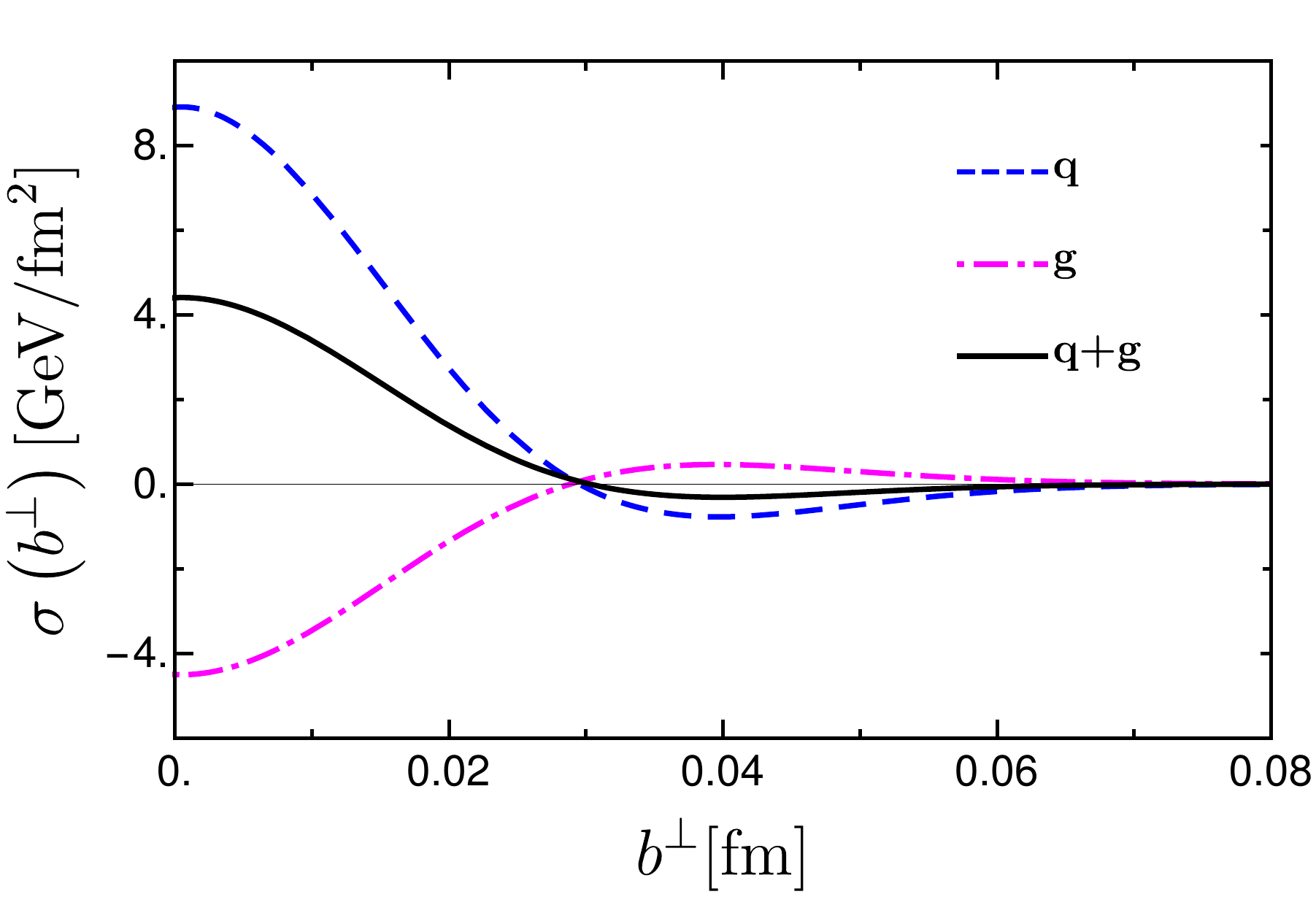}
\end{minipage}
\begin{minipage}{0.45\linewidth}
\includegraphics[scale=0.4]{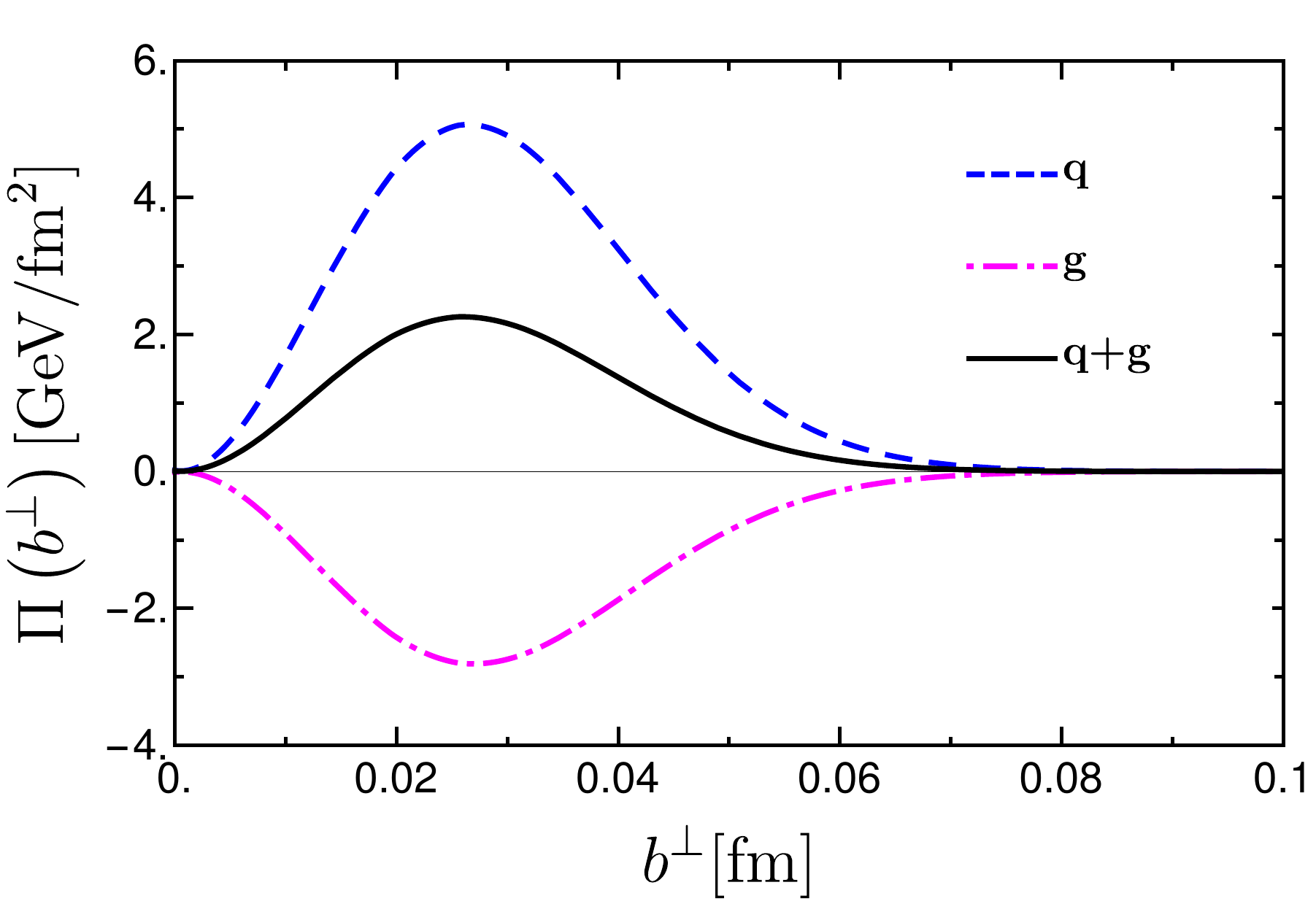}
\end{minipage}  
\caption{Plots of the 2D isotropic pressure $\sigma(b^{\perp})$ and pressure anisotropy $\Pi(b^{\perp})$ as a function of $b^{\perp}$. The dashed blue curve and the dot-dashed magenta curve are for the quark ($q$) and gluon ($g$) contributions respectively. The solid black curve is for the sum of quark and gluon ($q+g$) contribution. Here $\Delta=0.2$.}
\label{iso_aniso}
\end{figure}

 Fig.~\ref{iso_aniso} shows the plot for the isotropic pressure and the pressure anisotropy as defined in Eq.~\ref{totalP} and Eq.~\ref{shearlike} respectively. These two pressures are linear combinations of the radial and tangential pressure as shown in Eq.~\ref{sigmapi}. The isotropic pressure is the average of the radial and tangential pressure. Hence we observe that the region near the center resembles the behavior as seen in the radial pressure and the region around the periphery mimics the behavior observed in the tangential pressure. 
 Pressure anisotropy is the difference between radial and tangential pressure. In Fig.~\ref{iso_aniso}, we observe that the quark contribution is positive and the gluon contribution is negative. This implies that the radial pressure is always greater than the tangential pressure for the quark and for the gluon the tangential pressure is always greater than the radial pressure. The total pressure anisotropy remains non-negative throughout the impact parameter space.
		
\section{Conclusion} 
		
In this work, we have calculated  the gluon contribution to the GFFs and mechanical properties like pressure, shear force and energy distribution for a dressed quark state.  A lot of theoretical studies have been done in the direction of mechanical properties of the nucleons \cite{Lorce:2018egm,Goeke:2007fp,Neubelt:2019sou,Cebulla_2007,Chakrabarti:2015lba,Dorati_2008}, especially the \emph{D}-term has got a lot of attention because it is related to the pressure distribution inside the nucleon. Most phenomenological models for the nucleon do not include gluons. But the \emph{D}-term and the pressure distribution depend on the quark-gluon interaction \cite{Freese:2021czn,Polyakov:2018zvc}, as they are related to the transverse components of the energy-momentum tensor. In this work, instead of a nucleon state, we use a simpler relativistic spin-$\frac{1}{2}$ composite state, namely a quark dressed with a gluon at one loop in QCD. The advantage is that in the light-front gauge, one can use the constraint equations to eliminate the constrained fields. We use the two-component formalism in light-front Hamiltonian perturbation theory. The quark state is expanded in Fock space in terms of multi-parton LFWFs and we keep up to the two-particle sector that is the quark-gluon LFWF. These Boost invariant LFWFs for a dressed quark state can be calculated analytically using the LFQCD Hamiltonian. This allows us the calculate the contribution to the GFFs coming from the quark and gluon parts of the energy-momentum tensor, as well as their contribution  to the mechanical properties like the pressure, shear and energy distributions. This gives an intuitive picture of the spatial distributions of the two-particle relativistic composite state. In our earlier publication\cite{More:2021stk}
we have calculated the GFFs and mechanical properties of the dressed quark state from the quark part of the EMT; in this work, we have presented the contribution from  the gluon part of the EMT.
We have compared our results with existing results in the literature for a dressed electron 
\cite{Metz:2021lqv}. Using the QED values of the parameters we obtain $D_{\gamma}(q^2)$, which diverges as $q^2 \rightarrow 0$, as a result, the total \emph{D}-term also diverges in this limit. This behaviour is discussed in the literature. We observe that in the GFFs $D(q^2)$ and $\overline{C}(q^2)$ contributions from the light front zero modes need to be carefully investigated.
		
\section*{Acknowledgments}
J. M.  would like to thank the Department of Science and Technology (DST), Government of India, for financial support through Grant No. SR/WOS-A/PM-6/2019(G). S. N. thanks the Chinese Academy of Sciences Presidents International Fellowship Initiative for the support via Grants No. 2021PM0021. A. M. would like to thank SERB-POWER Fellowship (SPF/2021/000102) for financial support. 

\appendix
\section{Extraction of $A_{\G}(q^2)$ and $B_{\G}(q^2)$:}
\label{appaAandB}

To extract GFFs $A_\G(q^2)$ and $B_\G(q^2)$ we use the $(+,+)$ component of the EMT. The operator structure of $\theta_{\G}^{++}$ in light-cone gauge $A^+=0$ is :
\begin{align}
\theta_{\G}^{++}=\left(F_a^{+i}\right)^2=\left(\partial^+A_a^i\right)^2.
\end{align}

While calculating all the GFFs, we categorize the contributions to the matrix elements as follows:
\begin{itemize}
    \item Single particle contribution with one quark in the initial as well as final state and we denote it by \textbf{1,D}.
    \item Non-diagonal contribution in which we calculate matrix element with one quark in the initial state and one quark and a gluon in the final state or vice-versa. This contribution is denoted by \textbf{ND}.
    \item The two-particle diagonal contribution in which we calculate the matrix element with one quark and a gluon in the initial state as well as in the final state and we denote it by \textbf{2,D}.
\end{itemize}

The diagonal one-particle sector does not contribute, as this would need $x=1$ for the gluon.
\begin{align}
\left[\mathcal{M}_{\uparrow\uparrow}^{++}+\mathcal{M}_{\downarrow\downarrow}^{++}\right]_{\text{1,D}}=&0,\\
\left[\mathcal{M}_{\uparrow\downarrow}^{++}+\mathcal{M}_{\downarrow\uparrow}^{++}\right]_{\text{1,D}}=&0.
\end{align}

The final expression for the matrix elements coming from the two-particle sector of $\theta_{\G}^{++}$ in terms of the overlap of the LFWFs is as follows: 
\be
\Big{\langle} P',S' \Big{|} \theta_{\G}^{++}  \Big{|} P,S \Big{\rangle}  &=& 2P^{+2} 
\sum_{\lambda^{'},\lambda , \sigma} \int  dx d^2\bska~ (2x)
\phi_{\sigma ,\lambda'}^{*S'}(1-x,-(\bska+(1-x)\bsq)) 
~\phi_{\sigma,\lambda}^{S}(1-x,-\bska).
 \ee
 
The diagonal contribution coming from the two-particle sector is given by

\be
\left[\mathcal{M}_{\uparrow\uparrow}^{++}+\mathcal{M}_{\downarrow\downarrow}^{++}\right]_{\text{2,D}}
\!\!\es\!\! 2 P^{+2} g^2C_F\int \!\!\! \left[x \bska\right]  \frac{\left[m^2(1-x)^4+(1+x^2)(\kappa^{\perp2} -x\bska \cdot \bsq)\right]}{D_1  \ D_2},~~~~~\\
\left[\mathcal{M}_{\uparrow\downarrow}^{++}+\mathcal{M}_{\downarrow\uparrow}^{++}\right]_{\text{2,D}}
\es \!\!\! 2P^{+2} g^2C_F\int \!\!\! \left[x \bska\right]\left(-iq^{(2)}\right) \frac{mx^2\left(1-x\right)^2}{D_1  \ D_2}.
\ee
The non-diagonal contribution coming from the overlap of one and two-particle sectors also vanishes
\begin{align}
\left[\mathcal{M}_{\uparrow\uparrow}^{++}+\mathcal{M}_{\downarrow\downarrow}^{++}\right]_{\text{ND}}=&0,\\
\left[\mathcal{M}_{\uparrow\downarrow}^{++}+\mathcal{M}_{\downarrow\uparrow}^{++}\right]_{\text{ND}}=&0.
\end{align}

Where 
\be
\left[x \bska\right] :\es \frac{dx ~d^2\bska }{8 \pi^3}.\\
D_1:\es\left[\kappa^{\perp2}+m^2x^2\right].\\
D_2:\es\left[\left(\bska+\left(1-x\right)\bsq \right)^2+m^2x^2\right].
\ee

 From Eq. $\ref{ag}$ we get 
\begin{align}
\lim_{q^2\to 0} A_\G(q^2) = \frac{g^2C_f}{2\pi^2}\left[\frac{5}{9}+\frac{1}{3}ln\left(\frac{\Lambda^2}{m^2}\right)\right]
\label{aglimit}.
\end{align}
		
Now the expression for the quark part of the GFF $A_{\Q}(q^2)$ as discussed in our previous work \cite{More:2021stk}, in the limit $q^2 \rightarrow 0$, we get
\begin{align}
\lim_{q^2\to 0} A_\Q(q^2) =1- \frac{g^2C_f}{2\pi^2}\left[\frac{5}{9}+\frac{1}{3}ln\left(\frac{\Lambda^2}{m^2}\right)\right]
\label{aqlimit}.
\end{align}
So it is clear from Eqs. \ref{aglimit} and \ref{aqlimit} that the total quark and the gluon GFF $A(q^2)$ satisfies the sum rule as expected.
\begin{align}
\lim_{q^2\to 0} \left(A_\Q(q^2)+A_\G(q^2)\right)=1.
\end{align}
Similarly from Eqs. \ref{bg} and Ref. \cite{More:2021stk} in the limit $q^2 \rightarrow 0$, we get	
\begin{align}
\lim_{q^2\to 0} B_\G(q^2) = -\frac{ g^2 C_f}{12\pi^2}, ~~~ \lim_{q^2\to 0} B_\Q(q^2) = \frac{g^2 C_f}{12\pi^2}.
\end{align}
		
So, the total quark and gluon GFF $B(q^2)$ satisfies the sum rule as expected.
\begin{align}
\lim_{q^2\to 0} \left(B_\Q(q^2)+B_\G(q^2)\right)=0. \label{bqlimit}
\end{align}
Eq.~\ref{aqlimit} and \ref{bqlimit} imply that the total angular momentum of the dressed quark state  is $J(0)=\frac{1}{2}$ as per Ji's sum rule \cite{Ji:1996ek}.
		
\section{Extraction of \(C_{\G}(q^2)\) and \(\overline{C}_{\G}(q^2)\):}\label{appaDandCbar}
In light-front gauge $A^+ =0$, the calculation of the GFFs $C_\G(q^2)$ and $\overline{C}_\G(q^2)$ involve the 
 transverse component $A^{\perp}$ and the longitudinal component $A^-$ of the gauge field. The transverse component of the gauge field is the independent dynamical degree of freedom in LFQCD. However, $A^-$ is not a dynamical variable and can be eliminated using the equation of motion. Thus one obtains the following constraint equation :
\begin{align}
	\frac{1}{2}\left(\partial^+A^-_a\right)= \left(\partial^kA^k_a\right)+2g \frac{1}{\partial^+}\left(\xi^{\dagger}T^a\xi\right)+g f_{abc}\frac{1}{\partial^+}\left(A^k_b\partial^+A^k_c\right)\label{constrainteqn}.
\end{align}

It should also be noted that the last term in Eq.\ref{constrainteqn} does not contribute at $O(g^2)$ and hence can be neglected. The components of the EMT required for extracting $C_\G(q^2)$ and $\overline{C}_\G(q^2)$ are $\theta_{\G}^{ij}$ and $\theta_{\G}^{+j}$. These EMT components upto order $g$ can be written in terms of the gauge fields as shown below :

\begin{align}
   \theta_{\G}^{+j}
    =& -F^{+\lambda}_aF^j_{a\lambda}
    \\\nn&= \underbrace{\left(\partial^kA_a^k\right)\left(\partial^+A_a^j\right)+\left(\partial^+A_a^k\right)\left(\partial^jA_a^k-\partial^kA_a^j\right)}_\text{2,D}~+~\underbrace{2g\left(\partial^+A_a^j\right)\frac{1}{\partial^+}\left(\xi^{\dagger}T^a\xi\right)}_{\text{ND}}\\&+\underbrace{gf_{abc}\left(\partial^+A_a^j\right)\frac{1}{\partial^+}\left(A_b^k\partial^+A_c^k\right)-gf_{abc}\left(\partial^+A_a^k\right)A_b^jA_c^k}_{\text{Do not contribute}} \label{thetaplj},\\
\nn \theta_{\G}^{ij} 
   =& -F^{i\lambda}_aF^j_{a\lambda} -\frac{1}{4}\left(F_{\lambda\sigma a}\right)^2
  \\=&\theta_{I}^{ij}+\theta_{II}^{ij},
	\end{align}
where, 
\begin{align}
   \nn \theta_{I}^{ij} =& -\frac{1}{2}\left(\partial^+A_a^i\right)\left(\partial^-A_a^j\right)+\left(\partial^jA_a^i\right)\left(\partial^kA_a^k\right)
+\frac{1}{2}\left(\partial^iA_a^k-\partial^kA_a^i\right)\left(\partial^jA_a^k-\partial^kA_a^j\right)+\text{terms}\{i\leftrightarrow j\}\\\nonumber
& -\delta^{ij}\bigg[ \frac{1}{2}\left(\partial^kA_a^k\right)^2-\frac{1}{2}\left(\partial^+A_a^k\right)\left(\partial^-A_a^k\right)+\frac{1}{4}\left(\partial^kA_a^l-\partial^lA_a^k\right)^2\bigg]
\\&+2g\left(\partial^iA_a^j\right)\frac{1}{\partial^+}\left( \xi^{\dagger} T^a \xi\right)+\text{terms}\{i\leftrightarrow j\}
\label{thetaI},
\end{align}


\begin{align}
\nn\theta_{II}^{ij} =& \frac{g}{2}f_{abc}\left(\partial^jA_a^i\right)\frac{1}{\partial^+}\left(A_b^k\partial^+A_c^k\right)+\text{terms}\{i\leftrightarrow j\} + \frac{g}{2}f_{abc}\left(\partial^+A_a^i\right)\left(A_b^-A_c^j\right)+\frac{g}{2}f_{abc}\left(\partial^+A_a^j\right)\left(A_b^-A_c^i\right)\\\nn&-gf_{abc}\left(\partial^iA_a^k-\partial^kA_a^i\right)A_b^{j}A_c^{k} -gf_{abc}A_b^iA_c^k\left(\partial^jA_a^k-\partial^kA_a^j\right)
 -\delta^{ij}\bigg[\frac{g}{2}f_{abc}\left(\partial^+A_a^k\right)A_b^-A_c^k \\&-\frac{g}{4}f_{abc}\left(\partial^kA_a^l-\partial^lA_a^k\right)A_b^kA_c^l-\frac{g}{4}f_{abc}A_b^kA_c^l\left(\partial^kA_a^l-\partial^lA_a^k\right)-\frac{g}{2}f_{abc}\frac{1}{\partial^+}\left(A_b^k\partial^+A_c^k\right)\left(\partial^kA_a^k\right)\bigg].
\label{thetaII}
\end{align}
In our two-component approach, we have not included the contribution from the light-front zero modes that is $k^+=0$. We calculate the GFFs up to order $O(g^2)$ and 
the terms in Eq.~\ref {thetaII} and the second line of \ref{thetaplj} do not contribute to our calculations. The terms relevant to our calculations are given in Eq.~\ref {thetaI} and the first line of \ref{thetaplj}.

In order to compute the matrix element of terms in Eq.~\ref {thetaI} and \ref{thetaplj} we consider the generic operator for the diagonal contribution  
\be 
\mathcal{O}^{\mu \nu i j} = \left( \partial^\mu A^i \right) \left( \partial^{\nu} A^{j} \right),
\ee
such that components can be $\left(\mu,\nu\right) \equiv (+,-,1,2)$ and $\left(i,j,k,l\right) \equiv (1,2).$ By substituting the appropriate Lorentz index we can obtain each term in Eq.~\ref {thetaI} and \ref{thetaplj}. The final expression for the general matrix elements in terms of the overlap of the two-particle LFWFs using the above operator is as follows: 

\be
\nn\Big{\langle} P',S' \Big{|} \mathcal{O}^{\mu \nu i j}  \Big{|} P,S \Big{\rangle}  &=& 
\sum_{\lambda^{'},\lambda , \sigma} \int \frac{dx d^2\bska}{x}
\phi_{\sigma ,\lambda'}^{*S'}(1-x,-(\bska+(1-x)\bsq)) 
~\phi_{\sigma,\lambda}^{S}(1-x,-\bska)\\&\times& 
\left(\mathcal{R}^{\mu \nu}\, \epsilon_{\lambda}^{i} \epsilon_{\lambda'}^{j*}+
\mathcal{S}^{\mu \nu}\, \epsilon_{\lambda'}^{i*} \epsilon_{\lambda}^{j}\right).
\label{eqgen}
 \ee

The expressions for $\mathcal{R}^{\mu \nu}$ and $\mathcal{S}^{\mu \nu}$ are shown in Table.~\ref{Table}.
Consider the generic operators for non-diagonal contribution
\begin{align}
    \mathcal{O}_1^{ij} = 2g\left( \partial^i A^j \right) \frac{1}{\partial^+}\left( \xi^{\dagger} T^a \xi\right) , ~~~ \mathcal{O}_2^{+j} = 2g\left( \partial^+ A^j \right) \frac{1}{\partial^+}\left( \xi^{\dagger} T^a \xi\right) .
\end{align}
 So the expression for matrix elements for terms in Eq.~\ref{thetaI} and Eq.~\ref{thetaplj} in terms of overlap of the two-particle LFWFs is as follows:
 \newpage
\begin{align}
    \Big{\langle} P',S' \Big{|} \mathcal{O}_1^{ ij}  \Big{|} P,S \Big{\rangle} =& 
\sum_{\sigma,\lambda} \frac{2g}{\sqrt{2(2\pi)^3}} \int dx d^2\bska \bigg[\frac{\left(-\kappa^i \epsilon_{\lambda}^j\right)}{x^{3/2}} \left( \chi^{\dagger}_{S'} T^a \chi_{\sigma}\right)
\phi^{S}_{\sigma,\lambda}\left( 1-x,-\bska\right) \nn \\ +& \phi^{*S'}_{\sigma,\lambda}\left( 1-x,-\bska\right)\frac{\left(-\kappa^i -xq^i\right)\epsilon_{\lambda}^{j*}}{x^{3/2}} \left( \chi^{\dagger}_{\sigma} T^a \chi_{S}\right)\bigg], \\
\Big{\langle} P',S' \Big{|} \mathcal{O}_2^{ +j}  \Big{|} P,S \Big{\rangle} =& 
\sum_{\sigma,\lambda} \frac{2gP^+}{\sqrt{2(2\pi)^3}} \int dx d^2\bska \bigg[\left(\frac{-\epsilon_{\lambda}^j}{\sqrt{x}}\right) \left( \chi^{\dagger}_{S'} T^a \chi_{\sigma}\right)
\phi^{S}_{\sigma,\lambda}\left( 1-x,-\bska\right) \nn \\ +& \phi^{*S'}_{\sigma,\lambda}\left( 1-x,-\bska\right)\left(\frac{-\epsilon_{\lambda}^{j*}}{\sqrt{x}}\right) \left( \chi^{\dagger}_{\sigma} T^a \chi_{S}\right)\bigg].
\end{align}

\begingroup
\setlength{\tabcolsep}{10pt} 
\renewcommand{\arraystretch}{1.8} 
	\begin{table}[t]
  \begin{center}
    \begin{tabular}{|c|c|c|c|}
    \hline
    $\mu$ & $\nu$ & $\mathcal{R}^{\mu \nu}$ & $\mathcal{S}^{\mu \nu}$\\
      \hline
      $+$ & $-$ & $\left(\bska+\bsq\right)^2$
      & $\kappa^{\perp2}$\\
      \hline
      $+$ & $k$ & $xP^+\left(  \kappa^{k}+q^{k}\right)$
      &$xP^{+}\left( \kappa^{k}\right)$ \\
      \hline
      $k$ & $l$ & $\left( \kappa^{k} \right)\left(  \kappa^{l}+q^{l}\right)$
      & $\left(  \kappa^{k}+q^{k}\right)\left( \kappa^{l}\right)$\\ \hline
    \end{tabular}
    \caption{Expression for the tensors $\mathcal{R}$ and $\mathcal{S}$ appearing in Eq.~\ref{eqgen} for different components of 
    $\mu$ and $\nu$. Here $k$ and $l$ are transverse indices such that $(k,l) = (1,2)$.}
    \label{Table}
  \end{center}
\end{table}
\endgroup

 In order to extract GFF $\overline{C}_\G(q^2)$ we utilize the equation derived from the conservation of EMT as shown in Eq.~\ref{cbar1}

\begin{align}
q_{\mu}\theta^{\mu (1)}_{\G}
= \frac{q^{\perp2}}{2P^+}\theta^{+ 1}_{\G}- q^{(1)}\theta^{11}_{\G}- q^{(2)}\theta^{2 1}_{G}
\label{cbar}.
\end{align}

The conservation of EMT finally leads to Eq.~\ref{cbar1} which is used to extract $\overline{C}_\G(q^2)$. The L.H.S of Eq.~\ref{cbar3} will not receive any contribution from the single particle sector.
\begin{align}
\left[q_{\mu}\mathcal{M}^{\mu (1)}_{\uparrow \downarrow} + q_{\mu}\mathcal{M}^{\mu (1)}_{\downarrow \uparrow}\right]_{\text{1,D}}=0.
\end{align}

The diagonal and non-diagonal contributions to the L.H.S of Eq.~\ref{cbar3} is as follows

\begin{align}
\nonumber \left[q_{\mu}\mathcal{M}^{\mu (1)}_{\uparrow \downarrow} + q_{\mu}\mathcal{M}^{\mu (1)}_{\downarrow \uparrow}\right]_{\text{2,D}}=& ~ig^2C_F \int \!\!\!\left[x \bska\right]\frac{m\left(1-x\right)}{D_1 \ D_2}\big[\kappa^{(2)}q^{(1)}q^{\perp2}\left(1-x\right)-2\kappa^{(1)2}q^{(1)}q^{(2)}x+2\kappa^{(1)}\kappa^{(2)}\left(q^{\perp2}-xq^{(2)2}\right)\\&+\kappa^{(1)}q^{(2)}q^{\perp2}\left(1-x+x^2\right)\big], \\
\left[q_{\mu}\mathcal{M}^{\mu (1)}_{\uparrow \downarrow} + q_{\mu}\mathcal{M}^{\mu (1)}_{\downarrow \uparrow}\right]_{\text{ND}}=& ~i g^2 C_Fq^{(1)}q^{(2)} \int \left[x\bska\right]\frac{m x}{D_1}.
\end{align}

 The extraction of $C_\G(q^2)$ involves both $B_\G(q^2)$ and $\overline{C}_\G(q^2)$ as shown in Eq.~\ref{rhsC}. Again the single particle sector contribution to the L.H.S of Eq.~\ref{rhsC} is zero.
 
\begin{align}
    \left[\mathcal{M}^{11}_{\uparrow \downarrow} +
	\mathcal{M}^{22}_{\uparrow \downarrow} +\mathcal{M}^{11}_{\downarrow \uparrow} + \mathcal{M}^{22}_{\downarrow \uparrow}\right]_{\text{1,D}}=0.
\end{align}

The diagonal and non-diagonal contributions to the L.H.S of Eq.~\ref{rhsC} is as follows

\be
\!\!\!\!\left[\mathcal{M}^{11}_{\uparrow \downarrow} +\mathcal{M}^{22}_{\uparrow \downarrow} +\mathcal{M}^{11}_{\downarrow \uparrow} + \mathcal{M}^{22}_{\downarrow \uparrow}\right]_{\text{2,D}}
&=& 2 ig^2C_F \int \!\!\!\left[x \bska\right]\frac{m\left(1-x\right)}{D_1 \ D_2}\Big[\kappa^{\perp2} q^{(2)}-\kappa^{(2)}{q}^{\perp2}-2q^{(1)}\kappa^{(1)}\kappa^{(2)} 
 \nn \\ &+& x\left(\kappa^{(2)}q^{\perp2}+\kappa^{(1)2}q^{(2)}-\kappa^{(2)2}q^{(2)}\right)\Big], \nnn \\
 \left[\mathcal{M}^{11}_{\uparrow \downarrow} +
	\mathcal{M}^{22}_{\uparrow \downarrow} +\mathcal{M}^{11}_{\downarrow \uparrow} + \mathcal{M}^{22}_{\downarrow \uparrow}\right]_{\text{ND}} &=& -i g^2 C_Fq^{(2)} \int \left[x\bska\right]\frac{m x}{D_1}.
\ee

From Eq. \ref{cbarg} we get 
\begin{align}
\lim_{q^2\to 0} \overline{C}_{\G}(q^2) = \frac{g^2C_f}{24\pi^2}\left[-\frac{1}{3}+ln\left(\frac{\Lambda^2}{m^2}\right)\right]
\label{cbarglimit}.
\end{align}
The expression for the quark part of the GFF $\overline{C}_{\Q}(q^2)$ as discussed in our previous work \cite{More:2021stk}, in the limit $q^2 \rightarrow 0$, we get
\begin{align}
\lim_{q^2\to 0} \overline{C}_{\Q}(q^2) =\frac{g^2C_f}{24\pi^2}\left[\frac{1}{3}-ln\left(\frac{\Lambda^2}{m^2}\right)\right]
\label{cbarqlimit}.
\end{align}

So, the total quark and gluon GFF $\overline{C}(q^2)$ at $q^2 \rightarrow 0$ is
\begin{align}
\lim_{q^2\to 0} \left(\overline{C}_\Q(q^2)+\overline{C}_\G(q^2)\right)=0. \label{cbarsum}
\end{align}

\section{Integrals used to calculate GFFs}\label{appaint}
These integrals are used to calculate the analytical forms of the GFFs
\begin{align}
\int_{-\infty}^{\infty} d^2\kappa^{\perp} \frac{1}{D_1}  =&~ \pi \ln \left[1+\frac{\Lambda^2}{m^2x^2}\right],\\
\int_{-\infty}^{\infty}d^2\kappa^{\perp}\, \frac{1}{D_1\ D_2}=&~\frac{2\pi}{q^{\perp2}\left(1-x\right)^2}\frac{\tilde{f_2}}{\tilde{f_1}},\\
\int_{-\infty}^{\infty}d^2\kappa^{\perp}\, \frac{\kappa^{(i)}}{D_1\ D_2}=&~-\frac{\pi q^{(i)}}{q^{\perp2}\left(1-x\right)}\frac{\tilde{f_2}}{\tilde{f_1}},\\
\int_{-\infty}^{\infty}d^2\kappa^{\perp}\, \frac{\kappa^{(1)}\kappa^{(2)}}{D_1\ D_2}=&~\frac{\pi q^{(1)}q^{(2)}}{q^{\perp2}}\left[-1+\left(\frac{1+\tilde{f_1}^2}{2\tilde{f_1}}\right)\tilde{f_2}\right],\\
\int_{-\infty}^{\infty}d^2\kappa^{\perp}\, \frac{(\kappa^{(i)})^2}{D_1\ D_2}=&~\pi \left[-\frac1{2}\tilde{f_1} \tilde{f_2}+\frac{1}{2}+\frac{ (q^{(i)})^2}{q^{\perp2}}\left(-1+\left(\frac{1+\tilde{f_1}^2}{2\tilde{f_1}}\right)\tilde{f_2}\right)\right]+\frac{\pi}{2} \ln\left(\frac{\Lambda^2}{m^2x^2}\right).\nnn  
\end{align}

\be
&&\tilde{f_1}:= \sqrt{1+\frac{4m^2x^2}{q^{\perp2}\left(1-x\right)^2}}.\\
&&\tilde{f_2}:= \ln\left(\frac{1+\tilde{f_1}}{-1+\tilde{f_1}}\right).
\ee

\bibliographystyle{apsrev}
\bibliography{references}

\end{document}